\documentclass[journal]{IEEEtran}
\usepackage{tikz}
\usetikzlibrary{bayesnet}
\usetikzlibrary{calc,positioning}

\usepackage{comment}
\usepackage{amsmath}
\DeclareMathOperator{\erf}{erf}
\usepackage{cases}
\usepackage{amsfonts}
\usepackage{graphicx}
\usepackage[caption=false,font=footnotesize]{subfig}
\usepackage{mathrsfs}
\begin{document}

\title{The Information Rate of Fiber-Wireless Communication Systems Based on Photonic Generation of RF Signals}

\author{
Long~Huang,~and~Jianping~Yao,~\IEEEmembership{Fellow,~IEEE}

\thanks{Manuscript received XXXX, XXXX; revised XXXX, XXXX. This work was supported by the Fundamental Research Funds for the Central Universities (21625338) (\emph{Corresponding author: Long Huang}).}

\thanks{Long~Huang is with Guangdong Provincial Key
Laboratory of Optical Fiber Sensing and Communications, Institute of Photonics Technology, Jinan University, Guangzhou 511443, China (e-mail: longhuang@jnu.edu.cn).}
\thanks{Jianping Yao is with the Microwave Photonics Research Laboratory, School of Electrical Engineering and Computer Science, University of Ottawa, Ottawa, ON K1N 6N5, Canada (e-mail: jpyao@uottawa.ca).}
}


\maketitle
\begin{abstract}
High-capacity fiber-wireless communication systems operating at high frequencies increasingly rely on photonic generation of radio-frequency (RF) signals. In these systems, optical signals are transmitted over optical fibers and detected by photodetectors, where RF signals are generated at frequencies equal to the difference between the optical carrier frequencies. A major performance-limiting impairment is the phase noise of the generated RF signals, which originates from the phase noise of the optical sources. In this paper, we develop comprehensive probabilistic models for the two principal configurations of fiber–wireless communication systems employing photonic RF generation. Based on these models, we propose an efficient numerical framework for calculating the information rate (IR). Numerical simulations are performed to validate the efficiency of the proposed algorithm and the results provide design guidelines for high-performance fiber–wireless systems.
\end{abstract}


\IEEEpeerreviewmaketitle

\section{Introduction}
Fiber-wireless communications, which integrate wireless and optical access technologies, have attracted sustained research interest due to their inherent advantages including low loss, wide bandwidth, and immunity to electromagnetic interference (EMI) \cite{yu2018tutorial,tornatore2017fiber}. Microwave photonic techniques have emerged as a promising solution for the generation and transmission of radio-frequency (RF) signals, overcoming the limitations of traditional copper-based distribution systems such as bandwidth constraints and high transmission loss \cite{waterhouse2015realizing,yao2009microwave}. In photonic RF generation systems, two optical signals propagate through an optical fiber to photodetectors (PDs), where RF signals are generated at center frequencies equal to the frequency difference between the optical signals. The two optical signals may originate from either:
\begin{itemize}
\item Two free-running lasers, resulting in incoherent signals \cite{xu2025long,zhang20244,pang2011100}, or
\item Filtered comb lines from an optical frequency comb (OFC), producing coherent optical signals \cite{browning2018gain,tokizane2023terahertz,jia2025low}.
\end{itemize}
Although the OFC approach reduces phase noise at the cost of increased system complexity, both configurations suffer from the conversion of optical phase noise into RF phase noise, which significantly degrades transmission performance. Despite substantial analysis and experimental characterization of this effect \cite{tokizane2023terahertz,jia2025low,Wei,Shao,Delmade,Monroy}, a comprehensive theoretical framework for evaluating the IR under the influence of the phase noise of the photonically generated RF signal has yet to be developed.
\par
The main contributions of this paper are summarized as follows:

\begin{enumerate}
    \item We develop comprehensive probabilistic channel models for fiber--wireless communication systems employing photonic RF generation, including both the dual-laser (LD2) configuration with mutually incoherent optical sources and the OFC configuration with coherent optical sources. The proposed models explicitly incorporate the phase noise characteristics of the optical sources and the resulting RF phase noise.

    \item We propose an efficient numerical framework for calculating the information rate (IR) of fiber--wireless communication systems with phase-noise-induced channel memory. By employing composite trapezoidal-rule quadrature, the proposed method achieves improved convergence and higher numerical accuracy compared with existing auxiliary-channel-based approaches.

    \item We investigate the impact of key system parameters on the achievable IR, including laser linewidth, the relative time delay between optical signals, and line-of-sight multiple-input multiple-output (LOS-MIMO) channel configurations. The obtained results provide useful insights into the design and optimization of high-performance fiber--wireless communication systems based on photonic RF signal generation.
\end{enumerate}
\par The remainder of this paper is organized as follows. Section II presents the system models for the LD2 and OFC configurations. Section III develops the proposed IR calculation algorithm. Simulation results are presented in Section IV, followed by the conclusions in Section V.
\par Throughout this paper, the notation is defined as follows. $\mathcal{N}(0,\sigma^2)$ denotes the normal distribution with mean $0$ and variance $\sigma^2$, $\mathcal{CN}(0,2\sigma^2)$ denotes the complex normal distribution with mean $0$ and variance $2\sigma^2$. $|z|$ denotes the magnitude of a complex number $z$, while $\mathrm{arg}(z)$ denotes its phase. $\mathrm{Re}[z]$ and $\mathrm{Im}[z]$ denote the real and imaginary parts of $z$. $\dagger$ denotes the complex conjugation. $|\!||\!|$ denotes Euclidean norm of a vector. Furthermore, $\{x\}_{1:n}$ denotes the sequence $\{x_1,\ldots,x_n\}$. We also denote $[I(t)]_{\mathrm{AC}}=I(t)-\overline{I(t)}$, where $\overline{I(t)}$ denotes time average of $I(t)$. The upper case $P(.)$ denotes the probability mass function while the lower case $p(.)$ denotes the probability density function (PDF). Besides, the notation $\erf(\cdot)$ denotes error function. In addition, $\mathrm{Var}(\cdot)$ denotes variance and $\mathrm{Cov}(\cdot)$ denotes covariance. Finally, $\boldsymbol{v}^{(b)}$ denotes the $b$-th component of the vector $\boldsymbol{v}$, and $\tilde{\boldsymbol{v}} = \log (\boldsymbol{v})$ indicates that the logarithm is applied componentwise to the vector.
\section{System models}
\begin{figure}[h]
\centering
\subfloat[]{\includegraphics[width=0.8\linewidth]{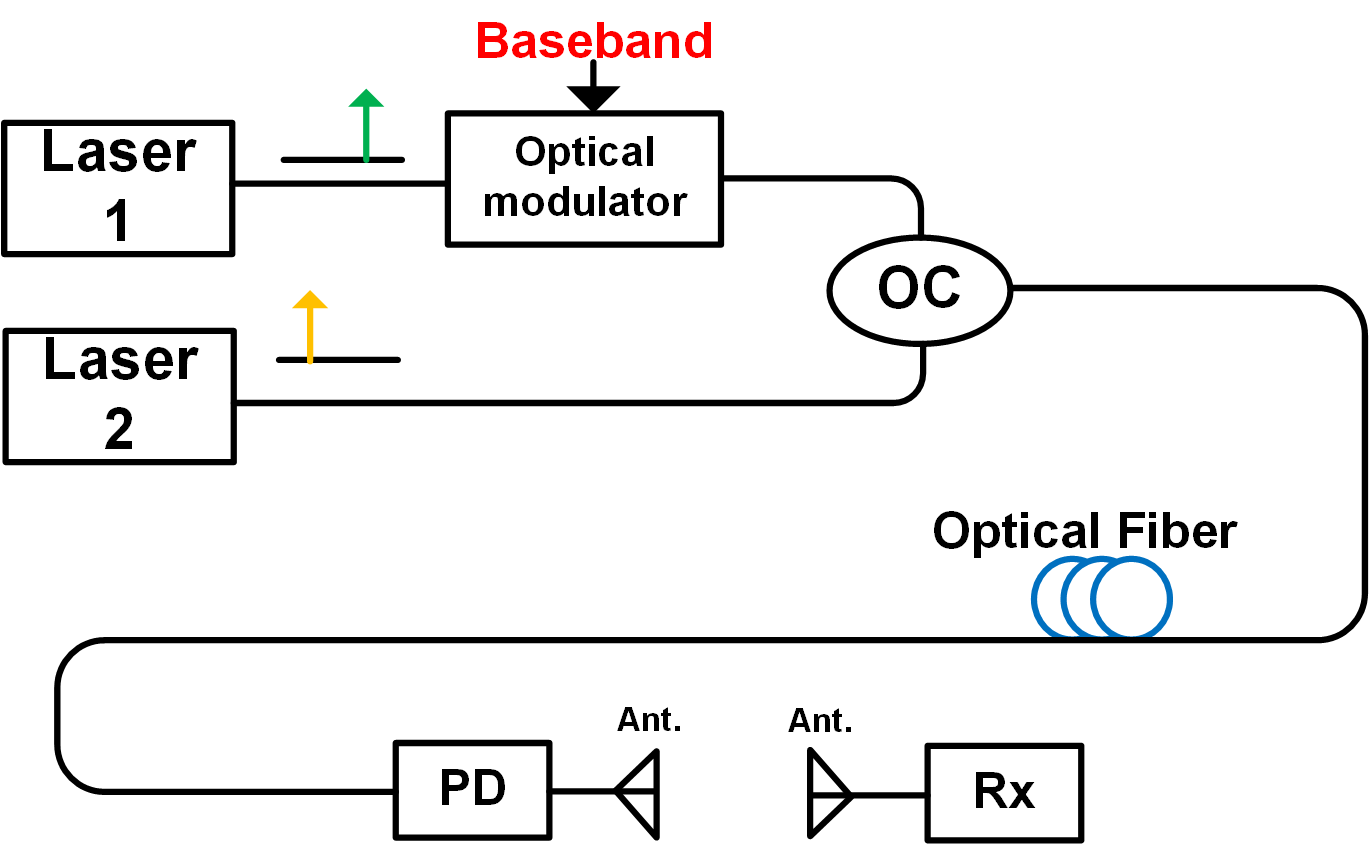}%
\label{config1_sisoI}}
\hfil
\subfloat[]{\includegraphics[width=0.8\linewidth]{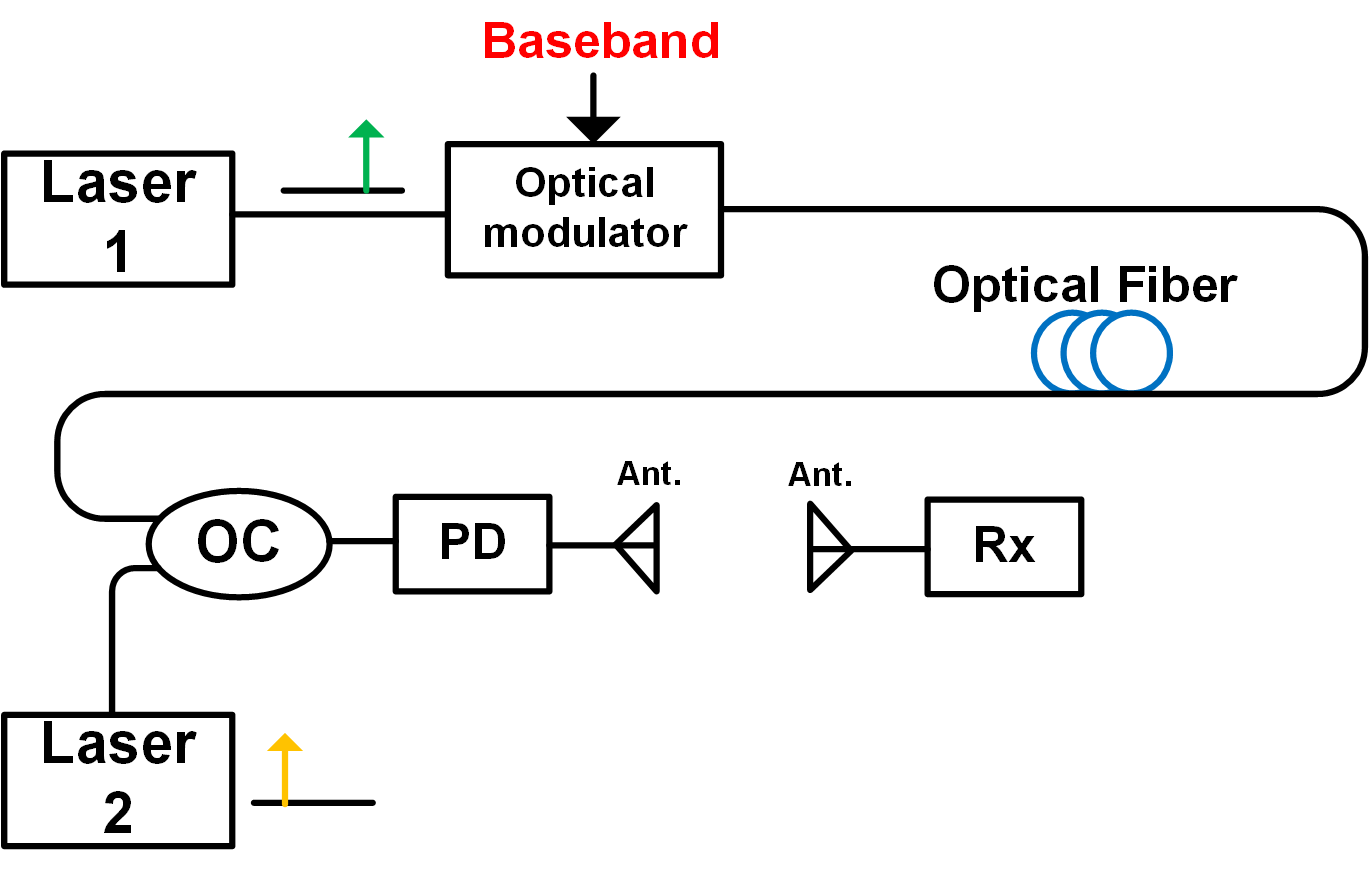}%
\label{config1_sisoII}}
\caption{Fiber-wireless communication systems based on photonic generation of RF signals, where the two optical signals are incoherent. (a) The two lasers are co-located, (b) the two lasers are not co-located.}
\label{config1_siso}
\end{figure}

\begin{figure}[!t]
    \centering
    \includegraphics[width=0.8\linewidth]{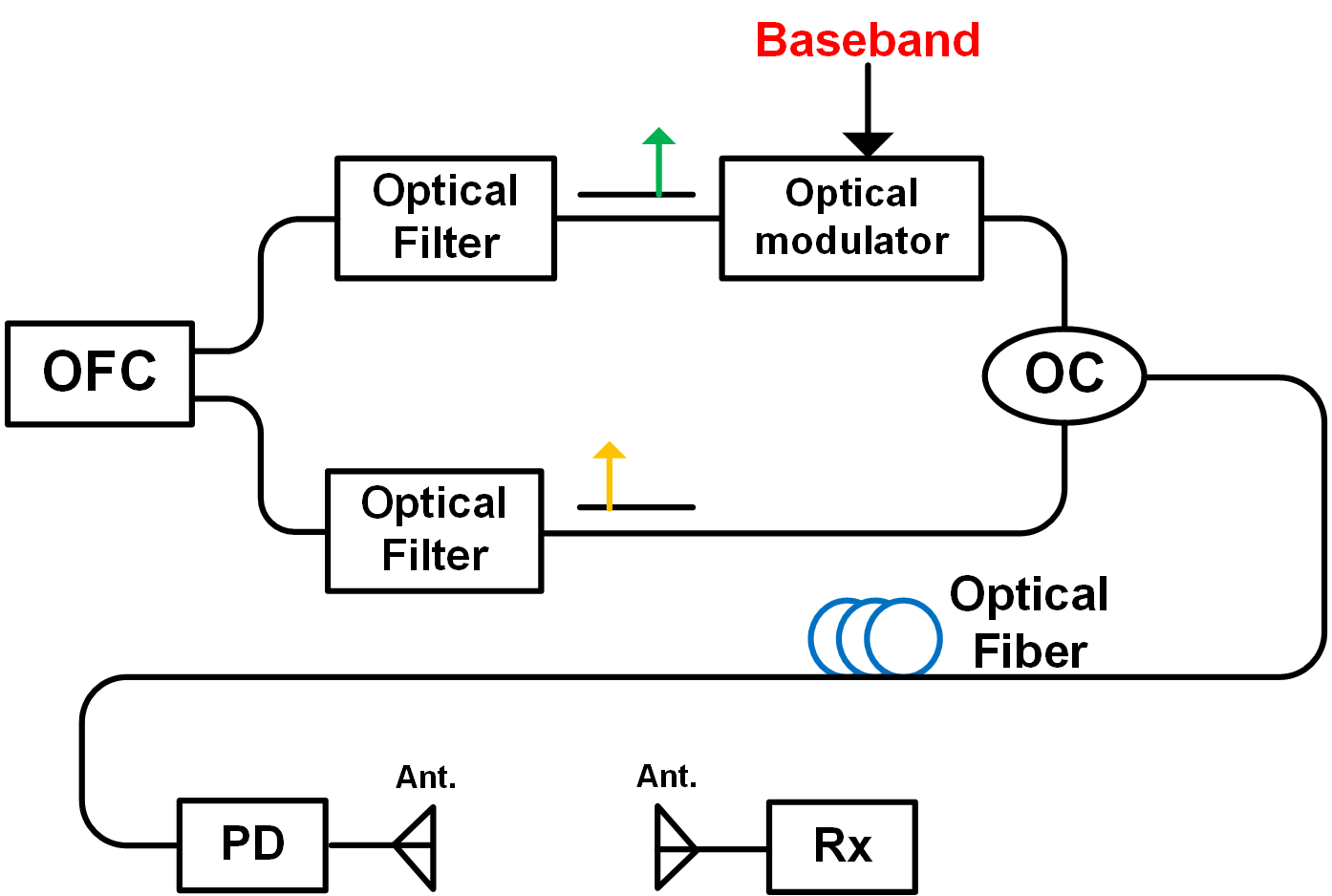}
    \caption{Fiber-wireless communication systems based on photonic generation of RF signals, where the two optical signals are coherent.}
    \label{config2}
\end{figure}

Two primary configurations are commonly employed in fiber–wireless communication systems for photonic RF signal generation. The first configuration, illustrated in Fig.~\ref{config1_siso}, employs two independent lasers. One optical carrier (from laser 1) is modulated by the baseband signal, while the second optical carrier (from laser 2) remains unmodulated. Depending on system deployment, the two lasers may either be co-located or not co-located. Since the optical signals originate from independent laser sources, they are mutually incoherent. This configuration is referred to as the LD2 configuration.
The second configuration, shown in Fig.~\ref{config2}, employs an OFC as the optical source. Two optical filters select different comb lines: one wavelength is modulated while the other remains unmodulated. The two optical signals are subsequently combined, transmitted through optical fiber, and converted into an RF signal by a PD. Since both wavelengths originate from a common optical source, they remain mutually coherent. This configuration is referred to as the OFC configuration.
\subsection{LD2 configuration}
In the LD2 configuration, two mutually incoherent optical signals generated by independent lasers are expressed as
\begin{subequations}
\begin{align}
{E_1}(t) &= {A_1}{e^{j\left[{{\omega_1}t+{\alpha_1}(t)} \right]}},\\
{E_2}(t) &= {A_2}(t){e^{j\left[ {{\omega_2}t+\theta(t)+{\alpha_2}(t)} \right]}},    
\end{align}    
\end{subequations}
where $A_1$, $\omega_{1}$ and $\alpha_{1}(t)$ are the amplitude, angular frequency, and phase noise of the first optical signal while $A_2(t)$, $\omega_{2}$, $\theta(t)$, and $\alpha_{2}(t)$ are the modulated amplitude, angular frequency, modulated phase and phase noise of the second optical signal. The phase noises are modeled as independent Wiener processes with increments satisfying 
\begin{subequations}
    \begin{align}
        {\alpha_1}(t +\Delta t) - {\alpha_1}(t) \sim \mathcal{N}(0,\sigma _{1,\alpha}^2(\Delta t)), \\
        {\alpha_2}(t + \Delta t) - {\alpha _2}(t) \sim \mathcal{N} (0,\sigma _{2,\alpha}^2(\Delta t)), 
    \end{align}
\end{subequations}
where the increment variances are related to the laser linewidths by
\begin{subequations}
    \begin{align}
        \sigma_{1,\alpha}^2(\Delta t) &= 2\pi {B_1}\Delta t,\\
        \sigma_{2,\alpha}^2(\Delta t) &= 2\pi {B_2}\Delta t,
    \end{align}
\end{subequations}
where $B_1$ and $B_2$ denote the full-width at half-maximum (FWHM) linewidths of the two lasers, respectively.\\
\indent The two optical signals are then sent to a PD for optical-to-electrical conversion. Assuming a relative time difference $\tau$ between the two signals, the generated RF signal is given by
\begin{IEEEeqnarray*} {l}
  {I_{RF}}(t) \propto {\left[ {{{\left| {{E_1}(t - \tau ) + {E_2}(t)} \right|}^2}} \right]_{\mathrm{AC}}} \\ 
   = {A_1}{A_2}(t)\cos\!\left[ {\left( {{\omega _2}\!-\!{\omega _1}} \right)t\!+\!\theta (t)\!+\!{\alpha _2}(t)\!-\!{\alpha _1}(t\!-\!\tau ) \!+\!{\omega _1}\tau } \right] \\ 
   = {A_1}{A_2}(t)\cos \left[ {{\omega _m}t + \theta (t) + \varphi (t) + {\omega _1}\tau } \right], \IEEEyesnumber
\end{IEEEeqnarray*}
where $\omega_{m}=\omega_{2}-\omega_{1}$ is the RF angular frequency, and 
\begin{align}
\varphi (t) = {\alpha _2}(t) - {\alpha _1}(t - \tau ) 
\end{align}
satisfies
\begin{IEEEeqnarray}{rCl}
\varphi (t + \Delta t) - \varphi (t) &\sim& \mathcal{N}(0,\sigma_\varphi^2),
\end{IEEEeqnarray}
where $\sigma_\varphi^2(\Delta t) = \sigma _{1,\alpha}^2(\Delta t) + \sigma _{2,\alpha}^2(\Delta t)$. Notably, the time delay $\tau$ only introduces a constant phase shift $\omega_1\tau$, which does not affect the statistical properties of the phase noise.\\
\indent Let $x(t)=A_2(t)e^{j\theta(t)}$ represent the input baseband signal sent to the channel, and express $I_{RF}(t)$ as $I_{RF}(t)=\mathrm{Re}[A_{1}e^{j{\omega_m}t}e^{j\omega_1{\tau}}y(t)]$, where $y(t)$ is the output baseband signal. The input-output relation is then $y(t) = x(t){e^{j\varphi (t)}}$. In the presence of additive white Gaussian noise (AWGN), the channel model becomes
\begin{align}
y(t) = x(t){e^{j\varphi(t)}} + n(t),
\end{align} 
where $n(t)$ is the AWGN term. Suppose a symbol interval of $T_0$, the discrete-time channel model is
\begin{IEEEeqnarray} {rCl}
{y_k} &=& {x_k}{e^{j{\varphi _k}}} + {n_k},
\end{IEEEeqnarray}
where $x_k,y_k, n_k,\varphi_k$ are samples of $x(t),y(t), n(t), \varphi(t)$, ${n_k} \sim \mathcal{CN}(0,2\sigma_n^2)$, $\varphi_{k}-\varphi_{k-1}\sim\mathcal{N}(0,\sigma_\varphi^2(T_0))$, and $\varphi_0$ follows uniform distribution within $[-\pi,\pi]$.\\ 
\begin{figure}[h]
    \centering
    \includegraphics[width=\linewidth]{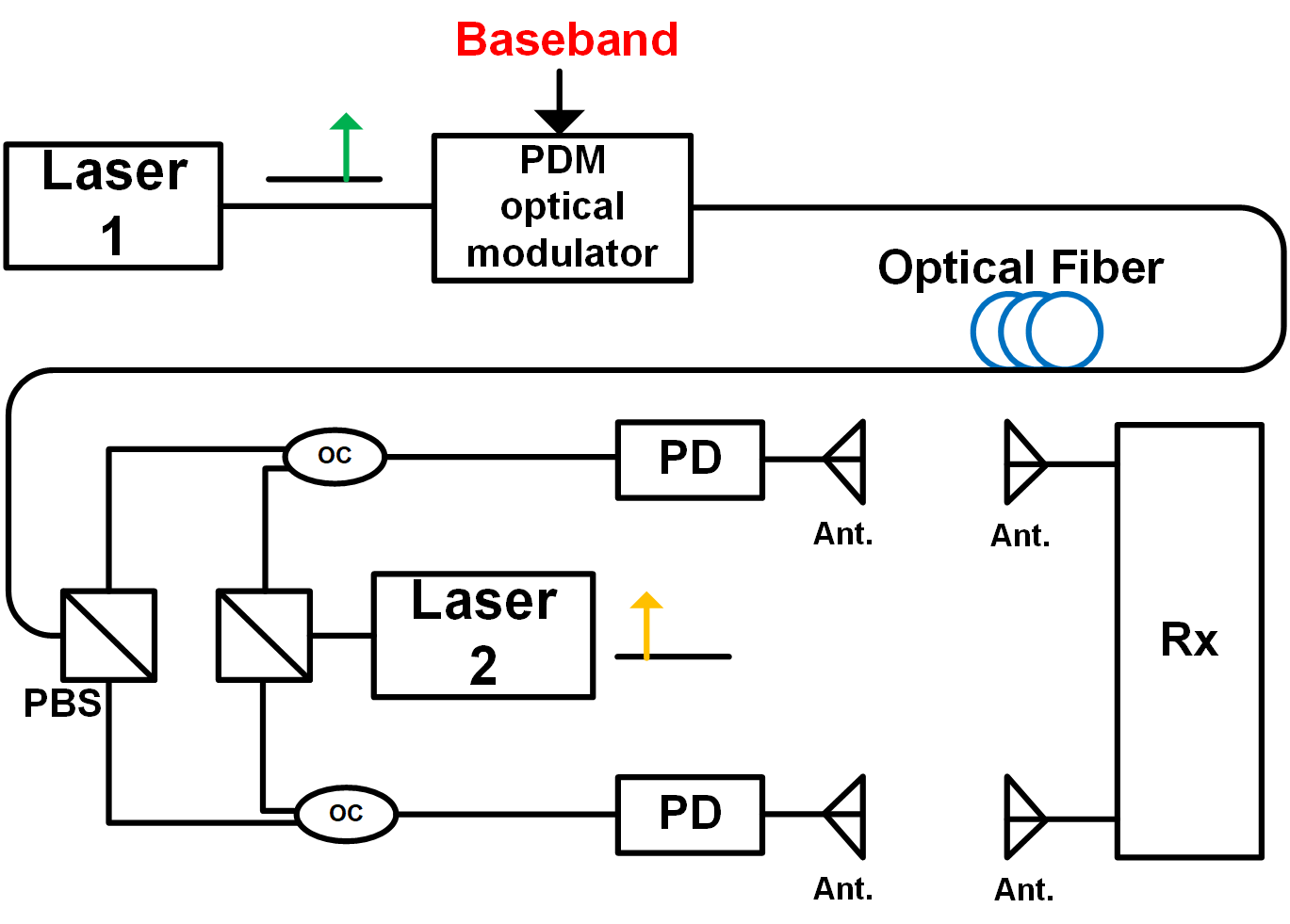}
    \caption{MIMO extension for the LD2 configuration.}
    \label{config1_mimo}
\end{figure}
\indent The above single-input single-output (SISO) model can be extended to MIMO systems \cite{pang2011100}. A representative MIMO implementation is illustrated in Fig.~\ref{config1_mimo}. At the optical baseband transmitter, the optical carrier generated by laser~1 is modulated using a polarization-division multiplexing (PDM) optical modulator to produce a PDM signal. After transmission through the optical fiber, the received optical PDM signal enters an optical heterodyne up-converter, where a polarization beam splitter (PBS) separates the X- and Y-polarization components. These optical signals are subsequently up-converted to the RF band using a second laser (laser~2). The generated RF signals are then radiated and received through antennas.

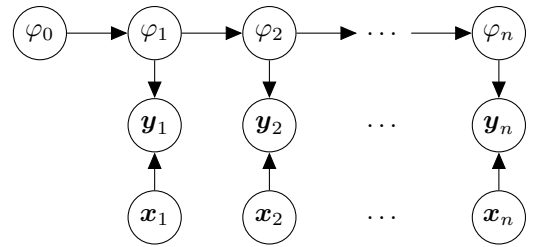
\begin{figure}[h]
    \centering
\begin{tikzpicture}
\node[latent] (f0) {$\varphi_0$};
\node[latent,right=0.8cm of f0] (f1) {$\varphi_1$};
\node[latent,right=0.8cm of f1] (f2) {$\varphi_2$};
\node [right=0.8cm of f2] (dots1) {$\dots$};
\node[latent,right=0.8cm of dots1] (fn) {$\varphi_n$};

\node[latent,below=0.5cm of f1] (y1) {$\boldsymbol{y}_1$};
\node[latent,below=0.5cm of f2] (y2) {$\boldsymbol{y}_2$};
\node [right=0.8cm of y2] (dots2) {$\dots$};
\node[latent,below=0.5cm of fn] (yn) {$\boldsymbol{y}_n$};

\node[latent,below=0.5cm of y1] (x1) {$\boldsymbol{x}_1$};
\node[latent,below=0.5cm of y2] (x2) {$\boldsymbol{x}_2$};
\node [right=0.8cm of x2] (dots3) {$\dots$};
\node[latent,below=0.5cm of yn] (xn) {$\boldsymbol{x}_n$};

\draw[->] (f0) -- (f1);
\draw[->] (f1) -- (f2);
\draw[->] (f2) -- (dots1);
\draw[->] (dots1) -- (fn);

\draw[->] (f1) -- (y1);
\draw[->] (f2) -- (y2);
\draw[->] (fn) -- (yn);
\draw[->] (x1) -- (y1);
\draw[->] (x2) -- (y2);
\draw[->] (xn) -- (yn);
\end{tikzpicture}
    \caption{A Bayesian network model for the LD2 configuration.}
    \label{fig:BN1}
\end{figure}

Assume that the numbers of transmit and receive antennas are $A$ and $B$, respectively. The MIMO channel matrix is denoted by $H\in\mathbb{C}^{B\times A}$, the transmitted signal by $\boldsymbol{x}_k\in\mathbb{C}^{A}$, the received signal by $\boldsymbol{y}_k\in\mathbb{C}^{B}$, and the noise signal by $\boldsymbol{n}_k\sim\mathcal{CN}(\boldsymbol{0},2\sigma_n^2\boldsymbol{I}_B)$. Accordingly, the discrete-time input-output relation is expressed as
\begin{IEEEeqnarray*} {l} \label{vec_sig_model}
\boldsymbol{y}_k = {H}\boldsymbol{x}_k{e^{j{\varphi _k}}} + \boldsymbol{n}_k 
= \boldsymbol{z}_k{e^{j{\varphi _k}}} + \boldsymbol{n}_k, \IEEEyesnumber
\end{IEEEeqnarray*}
where $\boldsymbol{z}_k=H\boldsymbol{x}_k$.\\
\indent Fiber-wireless communication systems operating at high RF frequencies typically use highly directive antennas and require line-of-sight (LOS) conditions. A MIMO system operating under these conditions is often referred to as a LOS-MIMO communication system. Since the $2\times 2$ configuration is the most widely adoptd architecture in fiber-wireless communication systems, as shown in Fig. \ref{config1_mimo}, we consider the $2\times 2$ LOS-MIMO channel as \cite{Bao:7248512}
\begin{align}
H = \begin{bmatrix}
e^{j\frac{2\pi}{\lambda}l_{11}} & e^{j\frac{2\pi}{\lambda}l_{12}}\\
e^{j\frac{2\pi}{\lambda}l_{21}} &e^{j\frac{2\pi}{\lambda}l_{22}}   
\end{bmatrix},    
\end{align}
where $l_{ij}$ denotes the path distance from the $i$-th transmitter antenna to the $j$-th receiver antenna.
For the simple case where $l_{11}=l_{22}=l_d$ and $l_{12}=l_{21}=l_x$, the LOS-MIMO channel simplifies to
\begin{align}
H = \begin{bmatrix}
1&e^{j\Delta}\\
e^{j\Delta}&1   
\end{bmatrix},   
\end{align}
with $\Delta = \frac{2\pi}{\lambda}(l_x-l_d)$.
The SISO channel can be seen as a special case of the MIMO channel with $A=1$, $B=1$, and $H=1$.\\
\indent Since $\varphi$ is the phase term which is wrapped within 2$\pi$ in nature, the transition PDF $p(\varphi_{k}|\varphi_{k-1})$ can be expressed as
\begin{align}\label{eqn:trans}
p\left( {{\varphi _k}|{\varphi_{k-1}}} \right) = \frac{1}{{\sqrt {2\pi \sigma _\varphi ^2} }}\sum\limits_{l =  - \infty }^\infty  {{e^{\frac{{ - {{\left( {{\varphi _k} - {\varphi_{k-1}} + 2\pi l} \right)}^2}}}{{2\sigma _\varphi ^2}}}}}.
\end{align}
The condtional PDF $p(\boldsymbol{y}_k|\boldsymbol{x}_k,\varphi_k)$ is given by
\begin{align}
p(\boldsymbol{y}_k|\boldsymbol{x}_k,\varphi_k) = \frac{1}{(2\pi\sigma_n^2)^B} \exp\bigg( -\frac{|\!|\boldsymbol{y}_k-H\boldsymbol{x}_k e^{j\varphi_k}|\!|^2}{2\sigma_n^2} \bigg),
\end{align}
where
\begin{align}
|\!|\boldsymbol{y}_k-H\boldsymbol{x}_k e^{j\varphi_k}|\!|^2 =|\!|\boldsymbol{y}_k-\boldsymbol{z}_k e^{j\varphi_k}|\!|^2= \sum_{b=1}^B |\boldsymbol{y}_k^{(b)}-e^{j\varphi_k}\boldsymbol{z}_k^{(b)}|^2.   
\end{align}
Finally, the channel input vector $\boldsymbol{x}$
is assumed to take values from the symbol alphabet $\mathcal{X}^A$.\\
\indent Consequently, the LD2 configuration of fiber-wireless communication systems forms a channel with memory, which can berepresented by the Bayesian network shown in Fig. \ref{fig:BN1}. \\

\subsection{OFC configuration}
In the OFC configuration, two mutually coherent optical signals generated from an optical frequency comb (OFC) are expressed as
\begin{IEEEeqnarray} {rCl} \label{part_coh}
\IEEEyesnumber 
\IEEEyessubnumber*
{E_1}(t) &=& {A_1}{e^{j\left( {{\omega _1}t + \beta (t)} \right)}},\\
{E_2}(t) &=& {A_2}(t){e^{j\left( {{\omega _2}t + \beta (t)} \right)}},
\end{IEEEeqnarray}
where $\beta(t)$ denotes the common Wiener phase noise process shared by the two optical signals. In Eq. (\ref{part_coh}), $\beta(t)$ is modeled as a Wiener phase noise satisfying
\begin{IEEEeqnarray}{rCl}
\beta(t+\Delta{t})-\beta(t) &\sim& \mathcal{N}(0,\sigma_{\xi}^2(\Delta{t})),
\end{IEEEeqnarray}
where the variance is related to the FWHM linewidth $B_l$ of the two coherent optical signals by
\begin{IEEEeqnarray} {rCl}
\sigma_\xi^2(\Delta t) &=& 2\pi {B_l}\Delta t.
\end{IEEEeqnarray}
In OFC-based fiber–wireless communication systems, the path difference at the transmitter, together with fiber chromatic dispersion introduce a relative delay between the two optical signals, commonly referred to as the \emph{walk-off effect} \cite{8543867}. The delay can be expressed as
\begin{align}
\tau = \tau_{d}+DL\cdot \Delta \lambda, 
\end{align}
where $\tau_d$ denotes the delay caused by the path difference at the transmitter, $D$ is the fiber chromatic dispersion coefficient, $L$ is the fiber length, and $\Delta\lambda$ is the wavelength spacing between the two optical signals. After photodetection, the generated RF signal is
\begin{IEEEeqnarray*} {l}
  {I_{RF}}(t) \\ 
   \propto {\left[ {{{\left| {{E_1}(t - \tau ) + {E_2}(t)} \right|}^2}} \right]_{\mathrm{AC}}} \\ 
   = {A_1}{A_2}(t)\cos \left[ {{\omega _m}t + \beta (t) - \beta (t - \tau ) + {\omega_2}\tau } \right] \\ 
   = {A_1}{A_2}(t)\cos \left[ {{\omega _m}t + \xi (t) + {\omega_2}\tau } \right], \IEEEyesnumber
\end{IEEEeqnarray*}
where 
\begin{align}
\xi(t) = \beta(t)-\beta(t-\tau)
\end{align}
is the time-delay-induced phase noise. Accordingly, the equivalent baseband input-output relation becomes
\begin{align}
y(t) = x(t){e^{j\xi (t)}}.
\end{align}
When AWGN is considered, the channel model becomes $y(t) = x(t)e^{j\phi(t)}e^{j\xi(t)} + n(t)$, where $n(t)$ is the AWGN term. Assuming a symbol interval of $T_0$, the corresponding discrete-time channel model is
\begin{IEEEeqnarray} {rCl}
{y_k} &=& {x_k}{e^{j{\xi _k}}} + {n_k},
\end{IEEEeqnarray}
where $y_k, x_k, \xi_k, n_k$ are samples of $y(t), x(t), \xi(t), n(t)$, ${n_k} \sim \mathcal{CN}(0,2\sigma _n^2)$.\\
\begin{figure}[h]
    \centering
    \includegraphics[width=\linewidth]{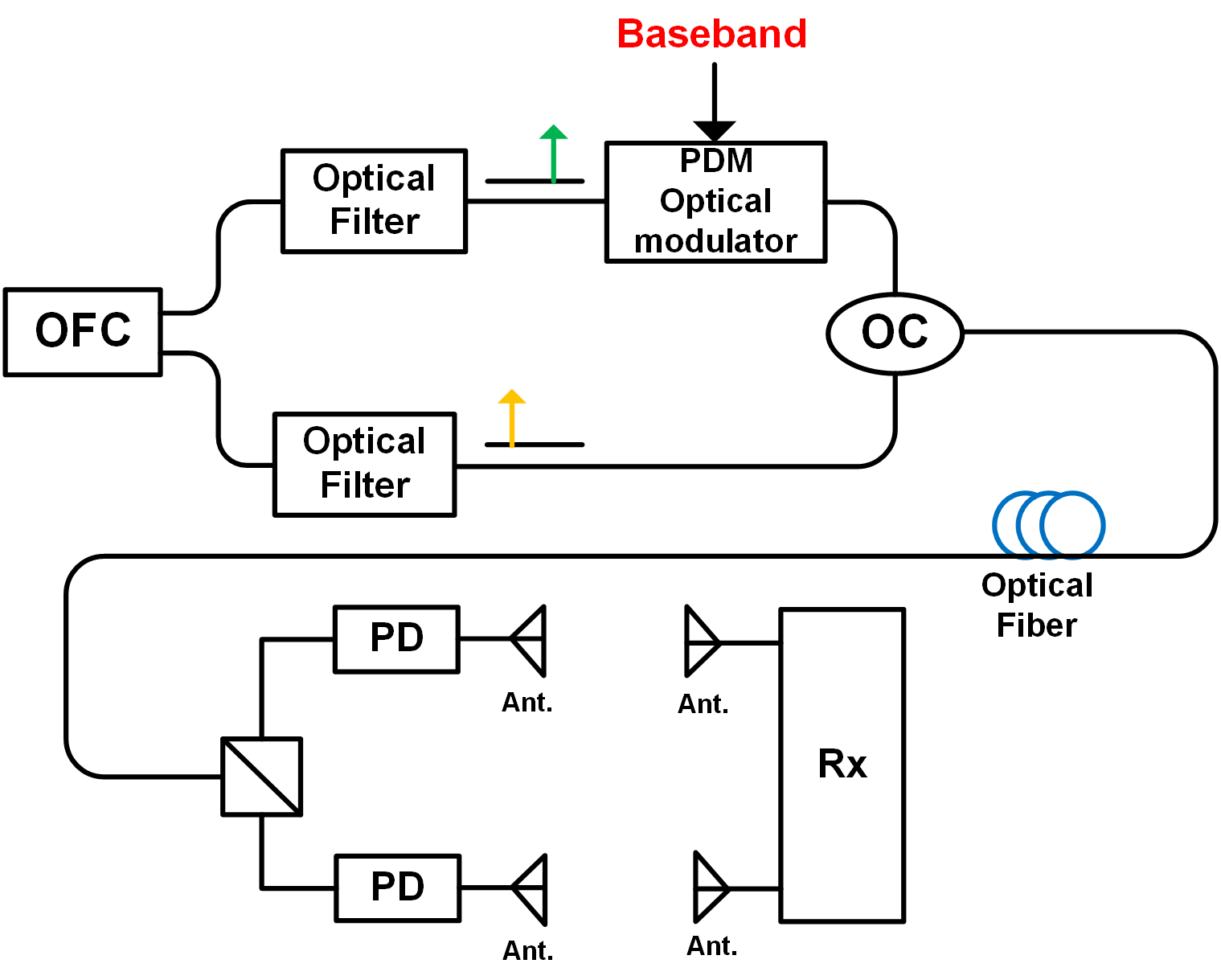}
    \caption{MIMO extension for the OFC configuration.}
    \label{config2_mimo}
\end{figure}
\indent The OFC configuration can also be extended to MIMO systems \cite{Kanno:12}. In a representative implementation, a PDM optical signal is generated using a PDM optical modulator, while an unmodulated optical carrier is aligned at an angle of $45^\circ$ relative to the X-polarization component. Both optical signals are transmitted through optical fiber and received by a polarization-diversity optical-to-electrical (O/E) converter. Within the O/E converter, the X- and Y-polarization components are separated and independently up-converted to the RF band. The resulting RF signals are subsequently radiated through  antennas. The discrete-time MIMO channel model is expressed as
\begin{IEEEeqnarray} {rCl}
\boldsymbol{y}_k &=& H\boldsymbol{x}_k{e^{j{\xi_k}}} + \boldsymbol{n}_k = \boldsymbol{z}_k{e^{j{\xi_k}}} + \boldsymbol{n}_k
\end{IEEEeqnarray}
where the MIMO channel is $H \in \mathbb{C}^{B \times A}$, the transmitter signal is $\boldsymbol{x}_k \in \mathbb{C}^A$, the receiver signal is $\boldsymbol{y}_k \in \mathbb{C}^B$, and the noise term $\boldsymbol{n}_k \in \mathbb{C}^B$. As shown in Fig. \ref{config2_mimo}, the most common MIMO configuration is the 2$\times$2 system, i.e. $A=2$, $B=2$ and $H\in\mathbb{C}^{2\times 2}$. The SISO channel can be seen as a special case of the MIMO channel where $A=1$, $B=1$, and $H=1$. \\
\indent The conditional PDF $p(\boldsymbol{y}_k|\boldsymbol{x}_k,\xi_k)$ of the system can be expressed as
\begin{align}
&p(\boldsymbol{y}_k|\boldsymbol{x}_k,\xi_k) = \frac{1}{(\pi2\sigma_n^2)^B}\exp\bigg( -\frac{|\!|\boldsymbol{y}_k-H\boldsymbol{x}_k e^{j\xi_k}|\!|^2}{2\sigma_n^2} \bigg).
\end{align}
The transition PDF $p(\xi_k|\xi_{k-1})$ of the system is given by
\begin{align}\label{eqn:trans_xi}
p(\xi_k|\xi_{k-1})=
\begin{cases}
p(\xi_k)=\frac{1}{{\sqrt {2\pi \sigma_{\xi} ^2}(\tau) }}\sum\limits_{l=-\infty}^\infty\!\! e^{ -\frac{(\xi_k+2\pi l)^2}{2\sigma_{\xi}^2(\tau)} }\!\!,\ \tau \leq T_0\\
\frac{1}{{\sqrt {2\pi \sigma_t ^2} }}\sum\limits_{l=-\infty}^\infty e^{ -\frac{(\xi_k-\kappa\xi_{k-1}+2\pi l)^2}{2\sigma_t^2} },\ \tau > T_0\\
\end{cases}    
\end{align}
where $\kappa=1-T_0/\tau$ and $\sigma_t^2=2\pi B_l T_0 (2-T_0/\tau)$. The derivation of Eq. (\ref{eqn:trans_xi}) is provided in Appendix A. The PDF of the initial phase noise $\xi_1$ is given by
\begin{align}
p(\xi_1) =   \frac{1}{{\sqrt {2\pi \sigma_\xi^2(T_0)} }}\sum\limits_{l=-\infty}^\infty{{e^{\frac{{-{{\left( {\xi_1 + 2\pi l} \right)}^2}}}{{2\sigma_\xi^2(T_0)}}}}}  
\end{align}
\indent The OFC channel model can therefore be represented by the Bayesian networks shown in Fig.~\ref{fig:BN2}. When $\tau\leq T_0$, $\xi_k$ are independent, resulting in a memoryless channel as shown in Fig. \ref{fig:BN2a}. When $\tau\leq T_0$, $\xi_k$ is dependent on $\xi_{k-1}$, resulting in a memory channel as shown in Fig. \ref{fig:BN2b}. 
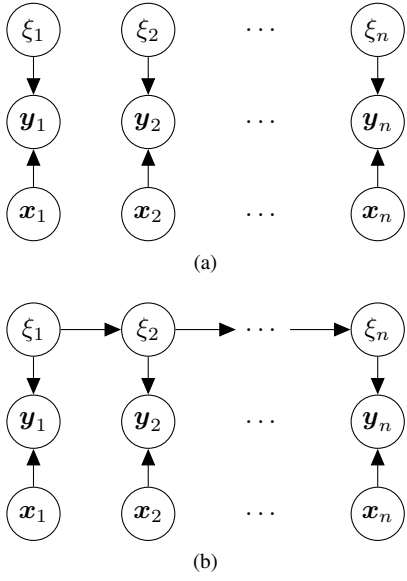
\begin{figure}[h]
\centering
\subfloat[]
{
\begin{tikzpicture}
\node[latent,right=0.8cm of f0] (f1) {$\xi_1$};
\node[latent,right=0.8cm of f1] (f2) {$\xi_2$};
\node [right=0.8cm of f2] (dots1) {$\dots$};
\node[latent,right=0.8cm of dots1] (fn) {$\xi_n$};

\node[latent,below=0.5cm of f1] (y1) {$\boldsymbol{y}_1$};
\node[latent,below=0.5cm of f2] (y2) {$\boldsymbol{y}_2$};
\node [right=0.8cm of y2] (dots2) {$\dots$};
\node[latent,below=0.5cm of fn] (yn) {$\boldsymbol{y}_n$};

\node[latent,below=0.5cm of y1] (x1) {$\boldsymbol{x}_1$};
\node[latent,below=0.5cm of y2] (x2) {$\boldsymbol{x}_2$};
\node [right=0.8cm of x2] (dots3) {$\dots$};
\node[latent,below=0.5cm of yn] (xn) {$\boldsymbol{x}_n$};


\draw[->] (f1) -- (y1);
\draw[->] (f2) -- (y2);
\draw[->] (fn) -- (yn);
\draw[->] (x1) -- (y1);
\draw[->] (x2) -- (y2);
\draw[->] (xn) -- (yn);
\end{tikzpicture}
\label{fig:BN2a}
}

\subfloat[]
{
\begin{tikzpicture}
\node[latent,right=0.8cm of f0] (f1) {$\xi_1$};
\node[latent,right=0.8cm of f1] (f2) {$\xi_2$};
\node [right=0.8cm of f2] (dots1) {$\dots$};
\node[latent,right=0.8cm of dots1] (fn) {$\xi_n$};

\node[latent,below=0.5cm of f1] (y1) {$\boldsymbol{y}_1$};
\node[latent,below=0.5cm of f2] (y2) {$\boldsymbol{y}_2$};
\node [right=0.8cm of y2] (dots2) {$\dots$};
\node[latent,below=0.5cm of fn] (yn) {$\boldsymbol{y}_n$};

\node[latent,below=0.5cm of y1] (x1) {$\boldsymbol{x}_1$};
\node[latent,below=0.5cm of y2] (x2) {$\boldsymbol{x}_2$};
\node [right=0.8cm of x2] (dots3) {$\dots$};
\node[latent,below=0.5cm of yn] (xn) {$\boldsymbol{x}_n$};

\draw[->] (f1) -- (f2);
\draw[->] (f2) -- (dots1);
\draw[->] (dots1) -- (fn);

\draw[->] (f1) -- (y1);
\draw[->] (f2) -- (y2);
\draw[->] (fn) -- (yn);
\draw[->] (x1) -- (y1);
\draw[->] (x2) -- (y2);
\draw[->] (xn) -- (yn);
\end{tikzpicture}
\label{fig:BN2b}
}

\caption{Bayesian network models for the OFC configuration: (a) $\tau\leq T_0$, (b) $\tau>T_0$.}
\label{fig:BN2}
\end{figure}

\section{IR Calculation}
To calculate IR for channels with memory, existing approaches in the literature primarily rely on particle filtering \cite{4418471} or auxiliary channel (AC) algorithms \cite{arnold2006simulation,barletta2012information}. Particle filtering typically suffers from slow convergence, whereas the AC algorithm is more efficient. In this work, we propose a novel algorithm based on composite trapezoidal-rule quadrature to calculate the IR, which achieves fast convergence and higher effciency than the conventional AC algorithm.\\
\indent Consider two discrete-time jointly stationary stochastic processes $\mathcal{X}$ as the input process and $\mathcal{Y}$ as the output process, taking values in discrete and continuous domains, respectively. The channel IR is defined as
\begin{IEEEeqnarray*} {rCl}
I &=& h(\mathcal{Y}) - h(\mathcal{Y}|\mathcal{X}) \\
&=& \lim_{n\rightarrow\infty} I_n\\
&=& \mathop {\lim }\limits_{n \to \infty } \left[ {{h_n}(\mathcal{Y}) - {h_n}(\mathcal{Y}|\mathcal{X})} \right], \IEEEyesnumber\\ 
\end{IEEEeqnarray*}
where $I_n={h_n}(\mathcal{Y})-{h_n}(\mathcal{Y}|\mathcal{X})$, ${h_n}(\mathcal{Y}) =  - {\log _2}(p({{\boldsymbol{y}}_{1:n}}))/n$ and ${h_n}(\mathcal{Y}|\mathcal{X}) = -{\log_2}(p({\boldsymbol{y}}_{1:n}|\boldsymbol{{x}}_{1:n}))/n$. In practice, two sufficiently long sequences of $\boldsymbol{x_{1:n}}$ and $\boldsymbol{y_{1:n}}$ are utilized, and then ${h_n}(\mathcal{Y})$ and ${h_n}(\mathcal{Y}|\mathcal{X})$ are utilized to compute $h_n(\mathcal{Y})$ and $h_n(\mathcal{Y|X})$, respectively, from which the IR is obtained.
\subsection{IR of LD2 configuration}
In this subsection, we consider the IR of the LD2 configuration.\\
\indent To calculate $h_n(\mathcal{Y}|\mathcal{X})$, it is necessary to evaluate $p(\boldsymbol{y}_{1:n}|\boldsymbol{x}_{1:n})$. Based on the probabilistic model shown in Fig. \ref{fig:BN1}, $p(\boldsymbol{y}_{1:n}|\boldsymbol{x}_{1:n})$ can be written as a multiple integral
\begin{align}
&p(\boldsymbol{y}_{1:n}|\boldsymbol{x}_{1:n}) \nonumber\\
&= \int_{-\pi}^\pi\cdots\int_{-\pi}^\pi p(\varphi_0)\prod_{i=1}^n p(\boldsymbol{y}_i|\boldsymbol{x}_i,\varphi_i)p(\varphi_i|\varphi_{i-1})d\varphi_{0:n}     
\end{align}
Let us denote 
\begin{IEEEeqnarray*} {rCl}
{\alpha_k}({\varphi_k}) &=& p(\boldsymbol{y}_{1:k},\varphi_k|\boldsymbol{x}_{1:k}). \yesnumber \\ 
\end{IEEEeqnarray*}
This leads to a recursive computation of $p(\boldsymbol{y}_{1:n}|\boldsymbol{x}_{1:n})$ as
\begin{align} \label{eqn:alpha}
\begin{cases}
{\alpha_0}({\varphi_0}) = p({\varphi_0}),\\
{\alpha_k}({\varphi_k}) = p(\boldsymbol{y}_k|\boldsymbol{x}_k,{\varphi_k})\!\!\int\limits_{ - \pi }^\pi\!{p({\varphi_k}|{\varphi_{k-1}}){\alpha_{k - 1}}\!({\varphi_{k - 1}})}d{\varphi_{k - 1}},\\
p(\boldsymbol{y}_{1:n}|\boldsymbol{x}_{1:n}) = \int\limits_{ - \pi }^\pi{{\alpha_n}({\varphi _n})d{\varphi_n}}.
\end{cases}
\end{align}
Since Eq. (\ref{eqn:alpha}) cannot, in general, be integrated analytically, we propose to employ the composite trapezoid-rule quadrature to approximate the integral in Eq. (\ref{eqn:alpha}). Suppose the integral interval $[-\pi,\pi]$ is equally divided into $T$ sub-intervals by $T+1$ points $-\pi=s_0,s_1,...,s_T=\pi$. Each sub-interval then has a length of $h=2\pi/T$, so $s_i=ih$, where $i=0,1,...,T$. Within each sub-interval $[s_i,s_{i+1}],i=0,1,...,T-1$, the trapezoid-rule quadrature is applied. Consequently, ${\alpha_k}({\varphi_k})$ is approximated by
\begin{align}
&{\alpha_k}({\varphi_k}) \notag\\
&= p(y_k|x_k,\varphi_k)\sum\limits_{i=0}^{T-1} {\bigg[{\int\limits_{{ih}}^{{(i+1)h}} {p({\varphi_k}|{\varphi_{k-1}})\alpha_{k-1}({\varphi_{k-1}})d{\varphi_{k-1}}}}\bigg]} \notag\\ 
&\approx p({{{y}}_k}|{x}_k,\varphi_k)\sum\limits_{i=0}^{T-1}
\bigg\{ \int\limits_{{ih}}^{{(i+1)h}} p({\varphi_k}|{\varphi_{k-1}})  \notag\\
& \left[{{\alpha_{k-1}({s_i})}{\ell_0}({\varphi_{k-1}})+\alpha_{k-1}({s_{i+1}}){\ell_1}({\varphi_{k-1}})} \right]d{\varphi_{k-1}} \bigg\}  \notag\\ 
&= p(y_k|x_k,\varphi_k)\sum\limits_{i = 0}^{T-1} {\left[ {{A_{i,0}}(\varphi_k)}\alpha_{k-1} ({s_{i}}) + {A_{i,1}(\varphi_k)}\alpha_{k-1} ({s_{i+1}}) \right]},  
\end{align}
where the weights $A_{i,j}$ are given by
\begin{IEEEeqnarray}{l} \label{A}
A_{i,j}(\varphi_k) = \!\!\!\!\int\limits_{{ih}}^{{(i+1)h}}\!\!\!\!{p({\varphi _k}|{\varphi _{k - 1}}){\ell_j}({\varphi _{k-1}})d{\varphi_{k-1}}},j=0,1,
\end{IEEEeqnarray}
with ${\ell_0}(\varphi_{k-1}) = \frac{\varphi_{k-1}-(i+1)h}{-h}$ and ${\ell_1}(\varphi_{k-1}) = \frac{\varphi_{k-1}-ih}{h}$ representing zero-order and first-order Lagrange polynomials, respectively. The weights can be expressed in closed form as
\begin{align}
A_{i,0} &= \frac{\sigma_\varphi}{\sqrt{2}h}\sum_{l=-\infty}^\infty \big[ \frac{e^{-v^2}}{\sqrt{\pi}}+v_1\mathrm{erf}(v)\big]\big\vert_{v=v_0}^{v_1}\\
A_{i,1} &= \frac{\sigma_\varphi}{\sqrt{2}h}\sum_{l=-\infty}^\infty \big[ \frac{e^{-v^2}}{\sqrt{\pi}}+v_0 \mathrm{erf}(v)\big]\big\vert_{v=v_1}^{v_0}     
\end{align}
where
\begin{align}
v_0 &= (\varphi_k-s_i+2\pi l)/(\sqrt{2}\sigma_\varphi),\notag\\
v_1 &= (\varphi_k-s_{i+1}+2\pi l)/(\sqrt{2}\sigma_\varphi).     
\end{align}
Consequently, $p(\boldsymbol{y}_{1:n}|\boldsymbol{x}_{1:n})$ can be approximated as
\begin{align}
\begin{cases}
{\alpha_0}({\varphi_0}) = p(\varphi_0),\\
{\alpha_k}({\varphi _k}) \approx p(\boldsymbol{y}_k|\boldsymbol{x}_k,{\varphi_k})\times\\
\ \ \sum\limits_{i = 0}^{T-1} [A_{i,0}(\varphi_k)\alpha_{k-1}\!(s_i)\!\!+\!\! A_{i,1}(\varphi_k)\alpha_{k-1}\!(s_{i+1})],\\
p(\boldsymbol{y}_{1:n}|\boldsymbol{x}_{1:n}) \approx \sum\limits_{i = 0}^{T - 1} {\left[ {\frac{{{h}}}{2}\alpha ({s_i}) + \frac{h}{2}\alpha ({s_{i + 1}})} \right]}.
\end{cases}
\end{align}
\indent Next, we calculate $p(\boldsymbol{y}_{1:n})$ to obtain $h_n(\mathcal{Y})$. Based on the probabilistic model shown in Fig. \ref{fig:BN1}, the term $p(\boldsymbol{y}_{1:n})$ can be expressed as a multiple integral
\begin{align}\label{eqn:py}
p(\boldsymbol{y}_{1:n}) = \int_{-\pi}^\pi\cdots\int_{-\pi}^\pi p(\varphi_0)\prod_{i=1}^n p(\varphi_i|\varphi_{i-1})p(\boldsymbol{y}_i|\varphi_i)d\varphi_{0:n}.
\end{align}
To calculate Eq. (\ref{eqn:py}), $p(\varphi_k|\varphi_{k-1})$
is given in Eq. (\ref{eqn:trans}) while $p(\boldsymbol{y}_k|\varphi_k)$ is obtained by marginalization of $p(\boldsymbol{y}_k|\boldsymbol{x}_k,\varphi_{k})$ as
\begin{IEEEeqnarray*} {rCl}
  p(\boldsymbol{y}_k|{\varphi_k}) &=& \sum\limits_{{\boldsymbol{x}_k} \in {\mathscr{X}^A}} {p(\boldsymbol{y}_k|\boldsymbol{x}_k,{\varphi_k})P(\boldsymbol{x}_k)},  \IEEEyesnumber\\
\end{IEEEeqnarray*}
Let us denote
\begin{IEEEeqnarray} {rCl}
{\beta_k}({\varphi_k}) &=& p({\boldsymbol{y}_{1:k}},\varphi_k).
\end{IEEEeqnarray}
Then, $p(\boldsymbol{y}_{1:n})$ can be calculated recursively as
\begin{align} \label{eqn:beta}
\begin{cases}
{\beta_0}({\varphi_0}) = p(\varphi_0),\\
{\beta_k}({\varphi_k})\!=\!p({\boldsymbol{y}_k}|{\varphi_k})\!\!\!\int\limits_{-\pi}^\pi\!{p({\varphi_k}|{\varphi_{k-1}}){\beta_{k-1}}({\varphi_{k-1}})}d{\varphi_{k-1}},\\
p({y_{1:n}}) = \int\limits_{-\pi}^\pi{{\beta_n}({\varphi_n})d{\varphi_n}}
\end{cases}.    
\end{align}
Similarily, by applying the trapezoid quadrature, Eq.(\ref{eqn:beta}) can be approximately calculated by
\begin{align}
\begin{cases}
{\beta_0}({\varphi_0}) &= p(\varphi_0),\\
{\beta_k}({\varphi _k}) &\approx p(\boldsymbol{y}_k|{\varphi _k})\sum\limits_{i = 0}^{T - 1}\big[A_{i,0}(\varphi_k)\beta(s_i) \\
&\quad + A_{i,1}(\varphi_k)\beta(s_{i+1})\big],\\
p(\boldsymbol{y}_{1:n}) &\approx \sum\limits_{i = 0}^{T - 1}\left[\frac{h}{2}\beta({s_i}) + \frac{h}{2}\beta({s_{i+1}})\right].
\end{cases}
\end{align}
To summarize, for calculating the IR, we introduce the matrices and vectors as
\begin{align} \label{notation}
&\boldsymbol{\alpha}_k=[\alpha_k(s_0),...,\alpha_k(s_T)]^T,\notag\\
&\boldsymbol{\beta}_k=[\beta_k(s_0),...,\beta_k(s_T)]^T,\notag\\
&\boldsymbol{h}=\frac{h}{2}[1,2,...,2,1]^T,\notag\\
&P_k = \operatorname{diag}(p(\boldsymbol{y}_k|\boldsymbol{x}_k,s_0),...,p(\boldsymbol{y}_k|\boldsymbol{x}_k,s_{T})),\notag\\
&P'_k = \operatorname{diag}(p(\boldsymbol{y}_k|s_0),...,p(\boldsymbol{y}_k|s_{T})),\notag\\
&C = 
\begin{bmatrix}
A_{0,0}(s_0),A_{0,1}(s_0)+A_{1,0}(s_0),\dots,A_{T-1,1}(s_0)\\
 A_{0,0}(s_1),A_{0,1}(s_1)+A_{1,0}(s_1),\dots,A_{T-1,1}(s_1)\\
 \vdots\\
A_{0,0}(s_T),A_{0,1}(s_T)+A_{1,0}(s_T),\dots,A_{T-1,1}(s_T)\\
\end{bmatrix}, \notag\\
&D_k = P_k C = \begin{bmatrix}
\boldsymbol{d}_{k,0}^T\\
\vdots\\
\boldsymbol{d}_{k,T}^T    
\end{bmatrix},\ D'_k = P_k' C = \begin{bmatrix}
\boldsymbol{d}_{k,0}^{'T}\\
\vdots\\
\boldsymbol{d}_{k,T}^{'T}  
\end{bmatrix}.
\end{align}
Then, we obtain the following matrix-form recursions for computing $p(\boldsymbol{y}_{1:n}|\boldsymbol{x}_{1:n})$ and $p(\boldsymbol{y}_{1:n})$
\begin{align} \label{eqn:config1}
 \begin{cases}
\boldsymbol{\alpha}_0 = [p(\varphi_0),...,p(\varphi_0)]^T,\\
\boldsymbol\alpha_k= D_k {\boldsymbol\alpha_{k-1}},\\
p(\boldsymbol{y}_{1:n}|\boldsymbol{x}_{1:n}) =\boldsymbol{h}^T\boldsymbol{\alpha}_n,\\
\boldsymbol{\beta}_0 = \boldsymbol{\alpha}_0,\\
\boldsymbol{\beta}_k=D'_k{\boldsymbol\beta_{k-1}},\\
p(\boldsymbol{y}_{1:n}) =\boldsymbol{h}^T\boldsymbol{\beta}_n. 
 \end{cases}   
\end{align}
After calculating $p(\boldsymbol{y}_{1:n}|\boldsymbol{x}_{1:n})$ and $p(\boldsymbol{y}_{1:n})$, the IR for a recursion length $n$ is given by
\begin{align} \label{eqn:IR}
I_n = \frac{-\log_2(p(\boldsymbol{y}_{1:n}))}{n} - \frac{-\log_2(p(\boldsymbol{y}_{1:n}|\boldsymbol{x}_{1:n}))}{n}.    
\end{align}
For sufficiently large $n$, $I_n$ converges to the true channel IR.

\indent To improve numerical stability and avoid arithmetic underflow during long recursive computations, all recursive operations are performed in the logarithmic domain. The log-domain version of the algorithm is given by
\begin{align} \label{eqn:log-version}
\begin{cases}
\tilde{\boldsymbol{\alpha}}_0 =  [\log_2{p(\varphi_0)},...,\log_2{p(\varphi_0)}]^T,\\
\tilde{\boldsymbol{\alpha}}_k^{(i)} = \log_2{\Big( \sum_{j=0}^{T} 2^{\big( \tilde{\boldsymbol{\alpha}}_{k-1}^{(j)} + \tilde{\boldsymbol{d}}_{k,i}^{(j)} \big)} \Big)},\ i=0,..,T,\\
\log_2(p(\boldsymbol{y}_{1:n}|\boldsymbol{x}_{1:n})) = \log_2{\Big( \sum_{j=0}^{T} 2^{\big( \tilde{\boldsymbol{\alpha}}_n^{(j)} + \tilde{\boldsymbol{h}}^{(j)} \big)} \Big)},\\
\tilde{\boldsymbol{\beta}}_0 = \tilde{\boldsymbol{\alpha}}_0, \\
\tilde{\boldsymbol{\beta}}_k^{(i)} = \log_2{\Big( \sum_{j=0}^{T} 2^{\big( \tilde{\boldsymbol{\beta}}_{k-1}^{(j)} + \tilde{\boldsymbol{d}}_{k,i}^{(j)} \big)} \Big)},\ i=0,..,T, \\
\log_2(p(\boldsymbol{y}_{1:n})) = \log_2{\Big( \sum_{j=0}^{T} 2^{\big( \tilde{\boldsymbol{\beta}}_n^{(j)} + \tilde{\boldsymbol{h}}^{(j)} \big)} \Big)},\\
I = \frac{-\log_2(p(\boldsymbol{y}_{1:n}))}{n} - \frac{-\log_2(p(\boldsymbol{y}_{1:n}|\boldsymbol{x}_{1:n}))}{n}.
\end{cases}
\end{align}
\subsection{IR of OFC configuration}
In this subsection, we consider the IR of the OFC configuration.\\
\indent When $\tau \leq T_0$, the channel model is shown in Fig. \ref{fig:BN2a}.
Based on the Bayesian network Fig. \ref{fig:BN2a}, the IR is calculated by
\begin{align}
I = -\log_2(p(\boldsymbol{y}_k)) - (-\log_2(p(\boldsymbol{y}_k \mid \boldsymbol{x}_k)))
\end{align}
where
\begin{align} \label{eq:Py_x}
&p(\boldsymbol{y}_k|\boldsymbol{x}_k) \notag\\
&=\int_{-\pi}^{\pi}
p(\boldsymbol{y}_k|\boldsymbol{x}_k,\xi_k)\,
p(\xi_k)\,d\xi_k
\notag\\
&=
\frac{
\exp\!\left(
-\dfrac{
\|\boldsymbol{y}_k\|^2+
\|H\boldsymbol{x}_k\|^2
}{2\sigma_n^2}
\right)
}{
(\pi 2\sigma_n^2)^B
}\times
\notag\\
&\quad
\Bigg[
I_0\!\left(
\frac{|z|}{\sigma_n^2}
\right)
+
2\sum_{m=1}^{\infty}
I_m\!\left(
\frac{|z|}{\sigma_n^2}
\right)
\exp\!\left(
-\frac{m^2\sigma_{\xi}^2(\tau)}{2}
\right)
\cos(m\phi)
\Bigg],
\end{align}
with
\begin{align}
w = \boldsymbol{x}_k^{\dagger}H^{\dagger}\boldsymbol{y}_k,\quad \phi = \arg(w)
\end{align}
and
\begin{IEEEeqnarray*} {rCl}
  p(\boldsymbol{y}_k|{\varphi_k}) &=& \sum\limits_{{\boldsymbol{x}_k} \in {\mathscr{X}^A}} {p(\boldsymbol{y}_k|\boldsymbol{x}_k,{\varphi_k})P(\boldsymbol{x}_k)},  \IEEEyesnumber\\ 
\end{IEEEeqnarray*}
The derivation of Eq. (\ref{eq:Py_x}) is given in Appendix B.\\
\indent When $\tau>T_0$, the system model is shown in Fig. \ref{fig:BN2b}. Based on the channel model presented in Fig. \ref{fig:BN2b}, $p(\boldsymbol{y}_{1:n} \mid \boldsymbol{x}_{1:n})$ and $p(\boldsymbol{y}_{1:n})$ are calculated by
\begin{align}
\begin{cases}
p(\boldsymbol{y}_{1:n}|\boldsymbol{x}_{1:n}) 
=\int_{-\pi}^\pi \cdots \int_{-\pi}^\pi \prod_{i=1}^n p(\boldsymbol{y}_i|\boldsymbol{x}_i,\xi_k)
p(\xi_1)\times \\
\quad\quad\quad\quad\quad\quad \prod_{i=2}^n p(\xi_i|\xi_{i-1})d\xi_{1:n}.\\
p(\boldsymbol{y}_{1:n}) =
\int_{-\pi}^\pi \cdots \int_{-\pi}^\pi
\prod_{i=1}^n p(\boldsymbol{y}_i|\xi_k)\times \\
\quad\quad\quad\quad\prod_{i=1}^n p(\xi_1) \prod_{i=2}^n p(\xi_i|\xi_{i-1})d\xi_{1:n}    
\end{cases}    
\end{align}
Similar to the LD configuration, if we define
\begin{align}
\begin{cases}
{\alpha_k}({\xi_k}) = p(\boldsymbol{y}_{1:k},\xi_k|\boldsymbol{x}_{1:k})\\
{\beta_k}({\xi_k}) = p({\boldsymbol{y}_{1:k}},\xi_k)    
\end{cases}    
\end{align}
the recursive computation is given by
\begin{align}
\begin{cases}
\alpha_1(\xi_1) = p(\boldsymbol{y}_1,\xi_1|\boldsymbol{x}_1)=p(\xi_1)p(\boldsymbol{y}_1|\boldsymbol{x}_1,\xi_1), \\
\alpha_k(\xi_k) = p(\boldsymbol{y}_k|\boldsymbol{x}_k,\xi_k)
\int_{-\pi}^{\pi} p(\xi_k|\xi_{k-1}) \,
\alpha_{k-1}(\xi_{k-1}) \, d\xi_{k-1}, \\
p(\boldsymbol{y}_{1:n}|\boldsymbol{x}_{1:n}) =
\int_{-\pi}^{\pi} \alpha_n(\xi_n) \, d\xi_n, \\
\beta_1(\xi_1) = p(\boldsymbol{y}_1,\xi_1), \\
\beta_k(\xi_k) = p(\boldsymbol{y}_k|\varphi_k)
\int_{-\pi}^{\pi} p(\xi_k|\xi_{k-1}) \,
\beta_{k-1}(\xi_{k-1}) \, d\xi_{k-1}, \\
p(\boldsymbol{y}_{1:n}) =
\int_{-\pi}^{\pi} \beta_n(\xi_n) \, d\xi_n.
\end{cases}
\end{align}
Then, the recursion can be approximately calculated using the trapezoidal-rule quadrature as
\begin{align}
\begin{cases}
{\alpha_1}({\xi_1}) = p(\xi_1)p(\boldsymbol{y}_1|\boldsymbol{x}_1,\xi_1),\\
{\alpha_k}({\xi_k}) \approx p(\boldsymbol{y}_k|\boldsymbol{x}_k,{\xi_k})\times\\
\ \ \sum\limits_{i = 0}^{T-1} [B_{i,0}(\xi_k)\alpha_{k-1}\!(s_i)\!\!+\!\! B_{i,1}(\xi_k)\alpha_{k-1}\!(s_{i+1})],\\
p(\boldsymbol{y}_{1:n}|\boldsymbol{x}_{1:n}) \approx \sum\limits_{i = 0}^{T - 1} {\left[ {\frac{{{h}}}{2}\alpha ({s_i}) + \frac{h}{2}\alpha ({s_{i + 1}})} \right]},\\
{\beta_1}({\xi_1}) = p(\xi_1)\sum_{\boldsymbol{x}_{1}\in\mathcal{X}^A} p(\boldsymbol{x}_1)p(\boldsymbol{y}_1|\boldsymbol{x}_1,\xi_1),\\
{\beta_k}({\varphi _k}) \approx p(\boldsymbol{y}_k|{\xi_k})\sum\limits_{i = 0}^{T - 1}\big[B_{i,0}(\xi_k)\beta(s_i) \\
\quad\quad\quad\quad + B_{i,1}(\xi_k)\beta(s_{i+1})\big],\\
p(\boldsymbol{y}_{1:n}) \approx \sum\limits_{i = 0}^{T - 1}\left[\frac{h}{2}\beta({s_i}) + \frac{h}{2}\beta({s_{i+1}})\right],
\end{cases}
\end{align}
where
\begin{align}
B_{i,j}(\xi_k) = \!\!\!\!\int\limits_{{ih}}^{{(i+1)h}}\!\!\!\!{p({\xi _k}|{\xi_{k - 1}}){\ell_j}({\xi_{k-1}})d{\xi_{k-1}}},j=0,1.   
\end{align}
The weights $B_{i,0}$ and $B_{i,1}$ admit the closed expressions
\begin{align}
B_{i,0} &= \frac{\sigma_t}{\sqrt{2}h\kappa^2}\sum_{l=-\infty}^\infty \big[ \frac{e^{-v^2}}{\sqrt{\pi}}+v_1\mathrm{erf}(v)\big]\big\vert_{v=v_0}^{v_1}\\
B_{i,1} &= \frac{\sigma_t}{\sqrt{2}h\kappa^2}\sum_{l=-\infty}^\infty \big[ \frac{e^{-v^2}}{\sqrt{\pi}}+v_0 \mathrm{erf}(v)\big]\big\vert_{v=v_1}^{v_0}     
\end{align}
where
\begin{align}
v_0 &= (\varphi_k-s_i+2\pi l)/(\sqrt{2}\sigma_t),\notag\\
v_1 &= (\varphi_k-s_{i+1}+2\pi l)/(\sqrt{2}\sigma_t).    
\end{align}
It is worth noting that
\begin{align}
\lim_{\tau\rightarrow\infty} B_{i,0} &= A_{i,0},\notag\\
\lim_{\tau\rightarrow\infty} B_{i,1} &= A_{i,1},   
\end{align}
which demonstrates that coherent optical signals gradually become incoherent as the time delay grows.\\
\indent By replacing the elements $A$ in $C$ by $B$ in Eq. (\ref{notation}), the log-domain version of the algorithm for the IR calculation of the OFC configuration is given by
\begin{align} \label{eqn:log-version}
\begin{cases}
\tilde{\boldsymbol{\alpha}}_1 =  [\log_2{p(\alpha_1(\xi_1))},...,\log_2{p(\alpha_1(\xi_1))}]^T,\\
\tilde{\boldsymbol{\alpha}}_k^{(i)} = \log_2{\Big( \sum_{j=0}^{T} 2^{\big( \tilde{\boldsymbol{\alpha}}_{k-1}^{(j)} + \tilde{\boldsymbol{d}}_{k,i}^{(j)} \big)} \Big)},\ i=0,..,T,\\
\log_2(p(\boldsymbol{y}_{1:n}|\boldsymbol{x}_{1:n})) = \log_2{\Big( \sum_{j=0}^{T} 2^{\big( \tilde{\boldsymbol{\alpha}}_n^{(j)} + \tilde{\boldsymbol{h}}^{(j)} \big)} \Big)},\\
\tilde{\boldsymbol{\beta}}_1 = \tilde{\boldsymbol{\alpha}}_1, \\
\tilde{\boldsymbol{\beta}}_k^{(i)} = \log_2{\Big( \sum_{j=0}^{T} 2^{\big( \tilde{\boldsymbol{\beta}}_{k-1}^{(j)} + \tilde{\boldsymbol{d}}_{k,i}^{(j)} \big)} \Big)},\ i=0,..,T, \\
\log_2(p(\boldsymbol{y}_{1:n})) = \log_2{\Big( \sum_{j=0}^{T} 2^{\big( \tilde{\boldsymbol{\beta}}_n^{(j)} + \tilde{\boldsymbol{h}}^{(j)} \big)} \Big)},\\
I = \frac{-\log_2(p(\boldsymbol{y}_{1:n}))}{n} - \frac{-\log_2(p(\boldsymbol{y}_{1:n}|\boldsymbol{x}_{1:n}))}{n}.
\end{cases}
\end{align}

\section{Simulation results}
\begin{figure}[!t]
    \centering
    \includegraphics[width=0.8\linewidth]{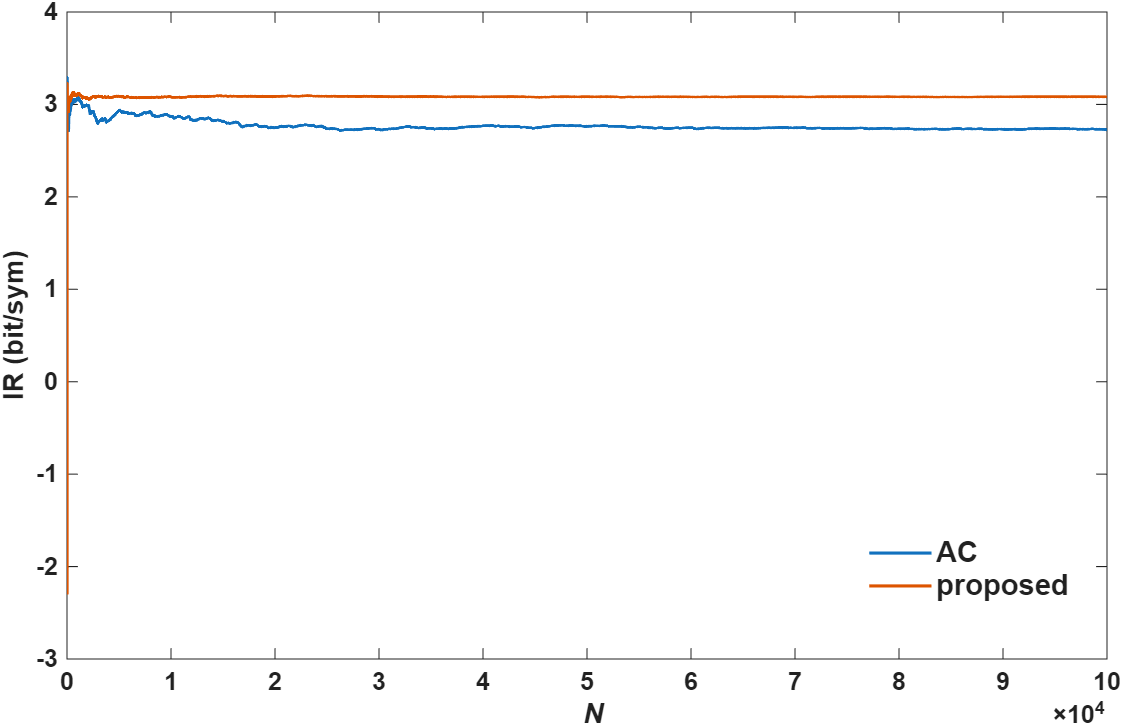}
    \caption{Calculated IR versus the recursion length for $T=20$ sub-intervals.}
    \label{fig:N1}
\end{figure}

\begin{figure}[!t]
    \centering
    \includegraphics[width=0.8\linewidth]{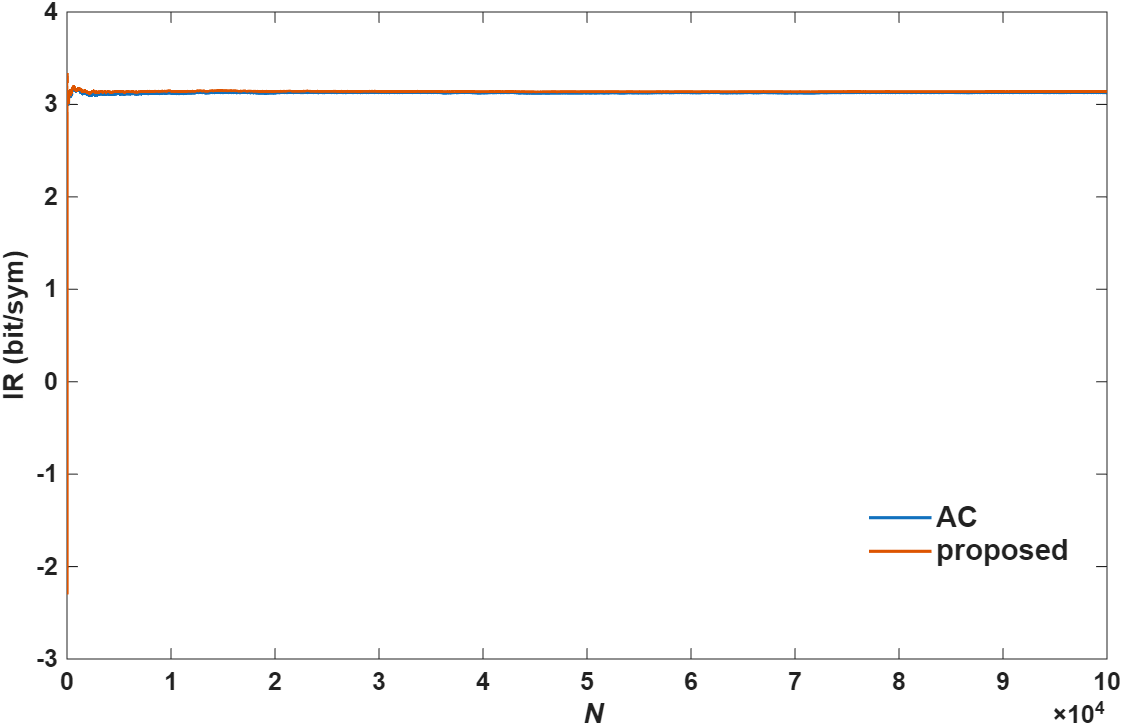}
    \caption{Calculated IR versus the recursion length for $T=150$ sub-intervals.}
    \label{fig:N2}
\end{figure}

\begin{figure}[!t]
    \centering
    \includegraphics[width=\linewidth]{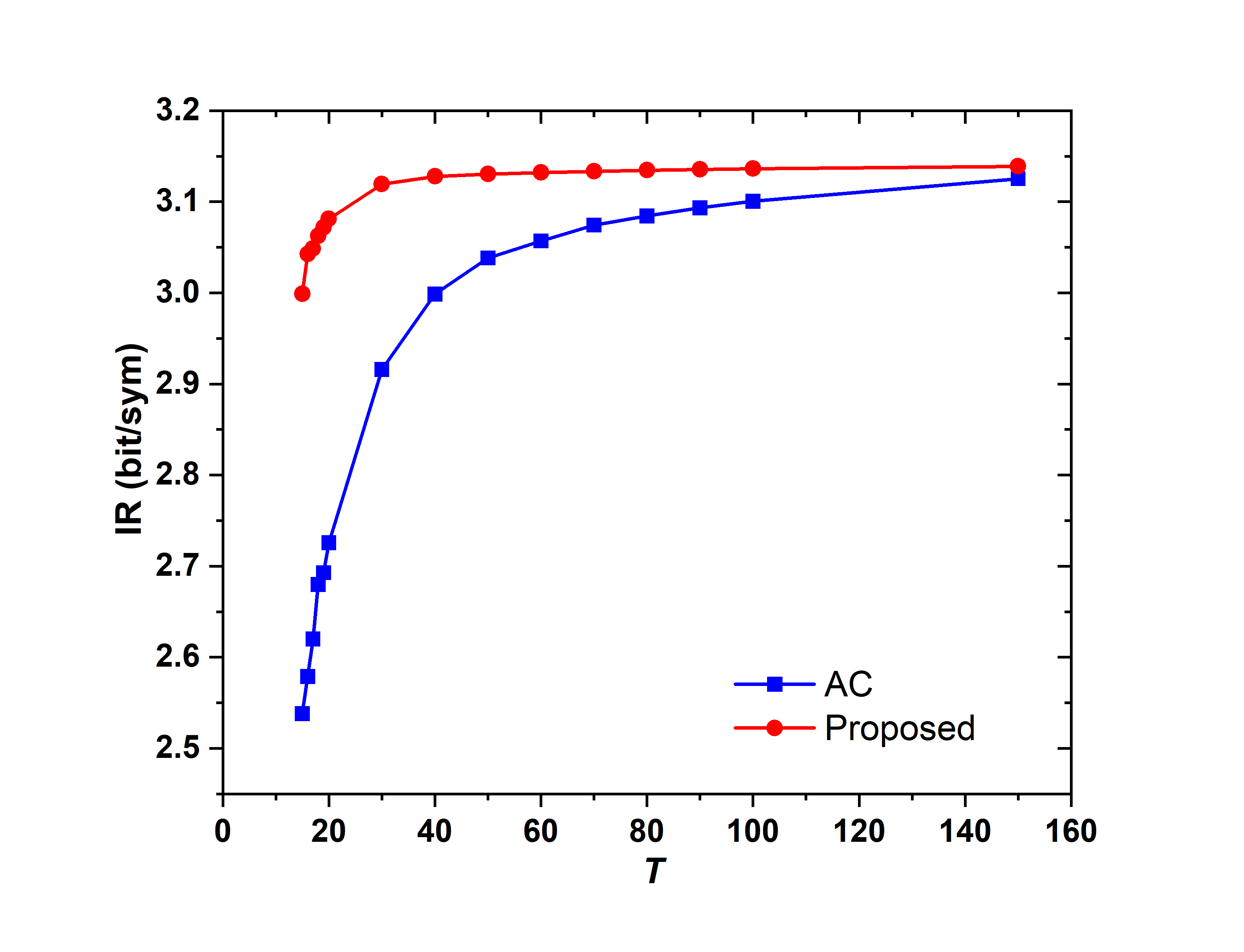}
    \caption{Calculated IR versus the number of sub-intervals $T$.}
    \label{fig:T}
\end{figure}

We compare the proposed algorithm with the AC algorithm from \cite{barletta2012information}. As a demonstration, the simulation considers the LD2 configuration with a laser linewidth of 200 kHz, which can be achieved using commercially available tunable laser sources reported in \cite{Lee}. The SNR is set to 10 dB. We first investigate the convergence behavior of the proposed algorithm. Fig. \ref{fig:N1} shows the calculated IR versus the sequence length when the number of sub-intervals is set to 20. We can see that the AC algorithm converges to 2.73 bit/sym while the proposed algorithm converges to 3.08 bit/sym. Then, the integral interval is divided into 150 sub-intervals. The calculated IR versus the sequence length is shown in Fig. \ref{fig:N2}. We can see that the AC algorithm converges to 3.13 bit/sym while the proposed algorithm converges to 3.14 bit/sym. Finally, the convergent values of the AC algorithm and the proposed algorithm versus the number of sub-intervals are calculated and the results are shown in Fig. \ref{fig:T}. As can be seen, the proposed algorithm converges faster than the AC algorithm because it introduces a smaller numerical integration error. The error analysis of both algorithms is provided in Appendix C. Owing to its higher efficiency, the proposed algorithm is adopted to evaluate the IR of fiber-wireless communication systems. In the following simulations, the number of sub-intervals is set to 100, the symbol rate is set to 20 GBd.

\begin{figure}[!t]
    \centering
    \includegraphics[width=\linewidth]{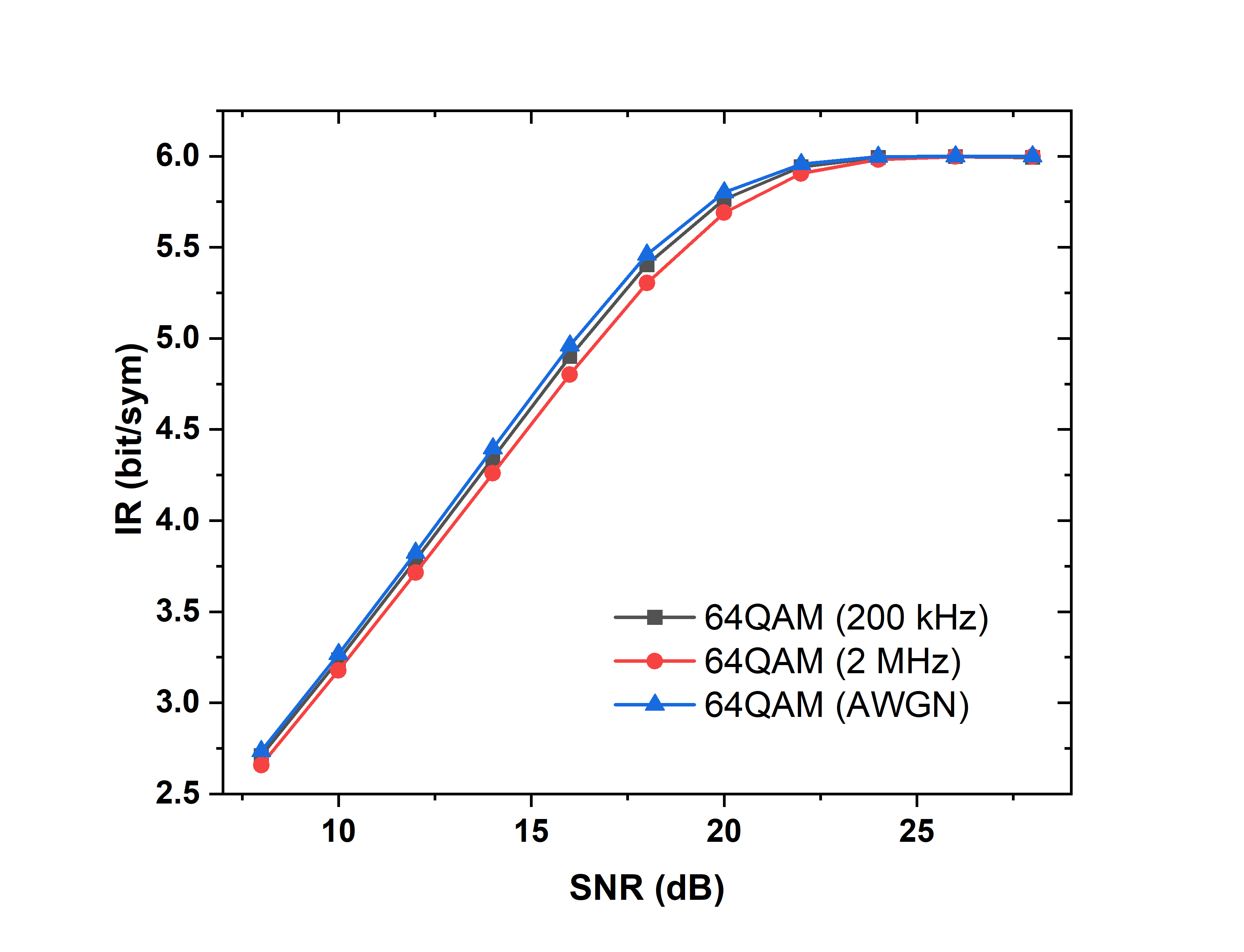}
    \caption{Calculated IR versus SNR for the LD2 configuration with different laser linewidths (64QAM input). The AWGN channle is shown for reference.}
    \label{fig:config1-qam64}
\end{figure}

\begin{figure}[!t]
    \centering
    \includegraphics[width=\linewidth]{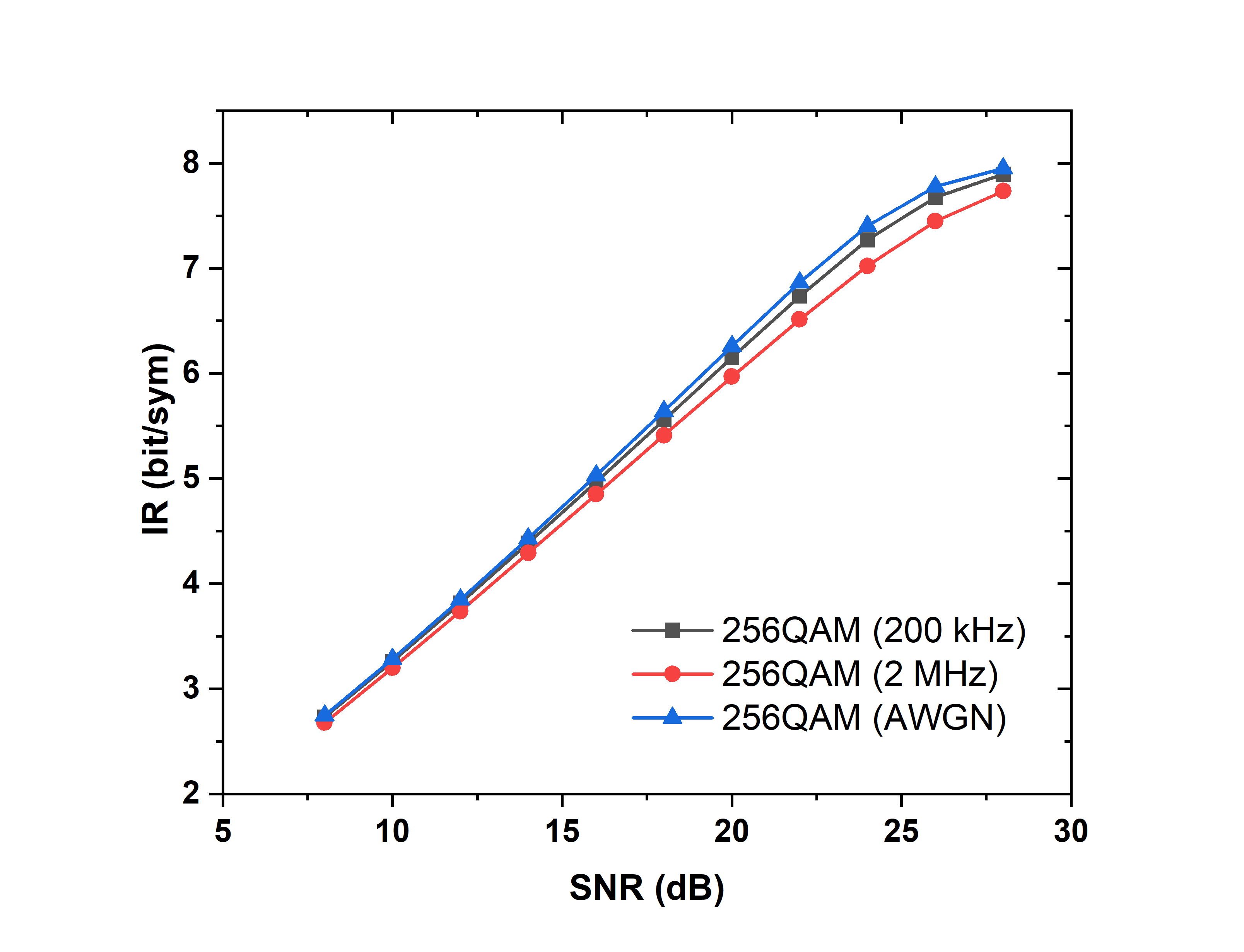}
    \caption{Calculated IR versus SNR for the LD2 configuration with different laser linewidths (256QAM input). The AWGN channel is shown for reference.}
    \label{fig:config1-qam256}
\end{figure}

\begin{figure}[!t]
    \centering
    \includegraphics[width=\linewidth]{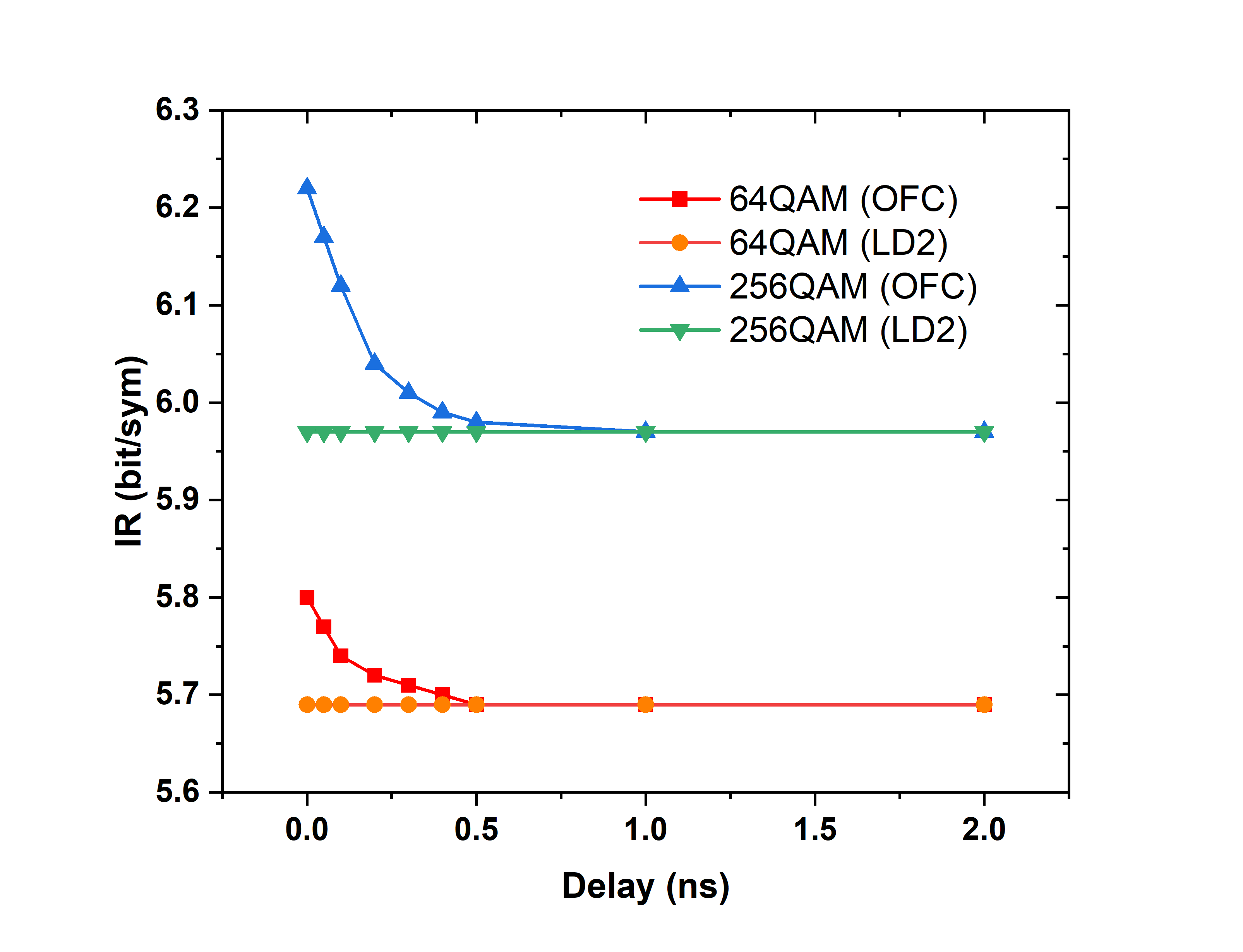}
    \caption{Calculated IR versus time delay for the OFC configuration (64QAM and 256QAM inputs). The LD2 configuration is shown for reference.}
    \label{fig:config2-delay}
\end{figure}

\indent Then we investigate the impact of key system parameters on the IR for fiber-wireless communication systems. First, we conduct simulations under the LD2 configuration to explore how the laser linewidth affects system performance. Fig. \ref{fig:config1-qam64} plots the IR against SNR for various laser linewidths. For comparison, we also simulate the ideal AWGN channel without phase noises. The results show that the IR rises with increasing SNR and decreasing the laser linewidth. At an SNR of 20 dB, for uniform 64QAM modulation, the IR reaches 5.76 bit/sym at a laser linewidth of 200 kHz, and drops to 5.69 bit/sym when the linewidth increases to 2 MHz. The corresponding IR for the ideal AWGN channel is 5.80 bit/sym. For uniform 256QAM under the same SNR condition, the IR is 6.15 bit/sym at 200 kHz and 5.97 bit/sym at 2 MHz, while the AWGN channel yields an IR of 6.26 bit/sym. As observed in Figs. \ref{fig:config1-qam64} and \ref{fig:config1-qam256}, despite a substantial rise in the laser linewidth from 200 kHz to 2 MHz, the resultant degradation in IR is relatively moderate.\\
\indent Then, the IR performance of the OFC configuration is simulated and compared with that of the LD2 configuration. In the simulation, the phase noise of the OFC is modeled with a linewidth of 2\,MHz. For reference, the LD2 configuration is simulated assuming each free-running laser also has a linewidth of 2.0\,MHz. The SNR is set to 20\,dB, and the time delay is varied from 0 to 2\,ns. The input signals are uniform 64QAM and 256QAM. The simulation results are shown in Fig.~\ref{fig:config2-delay}. It can be observed that the IR decreases as the time delay increases. For 64QAM, at a time delay of 0\,ns, the IR of the OFC configuration is 5.80\,bit/sym, while that of the LD2 configuration approaches 5.69\,bit/sym. As the time delay increases, the IR of the OFC configuration gradually declines, whereas the IR of the LD2 configuration remains unchanged. For instance, at a time delay of 0.5\,ns, the IR of the OFC configuration drops to 5.69\,bit/sym, which is nearly identical to that of the LD2 configuration. When the time delay $\tau$ exceeds 0.5\,ns, the IR saturates to the level of the LD2 configuration, as the two coherent lightwaves become effectively incoherent, yielding performance equivalent to that of the LD2 configuration. For 256QAM, at a time delay of 0\,ns, the IR of the OFC configuration is 6.22\,bit/sym, compared to 5.97\,bit/sym for the LD2 configuration. The OFC configuration outperforms the LD2 configuration by 0.25\,bit/sym, owing to the coherence between the two lightwaves in the OFC configuration, in contrast to the incoherence in the LD2 case. As the time delay increases, the IR of the OFC configuration again decreases, as the two coherent lightwaves progressively become incoherent. In contrast, the IR of the LD2 configuration remains constant. For example, at a time delay of 1\,ns, the IR of the OFC configuration falls to 5.97\,bit/sym, which is almost the same as that of the LD2 configuration, where the two lightwaves are completely incoherent. These results demonstrate that the performance advantage of the OFC configuration relies on preserving the mutual coherence between the optical carriers and gradually diminishes as the relative time delay increases.

Finally, we conduct simulations on the $2\times2$ line-of-sight (LOS) MIMO channel. Two distinct channel matrices are constructed by adjusting the phase difference between the direct and cross propagation paths. First, the inter-path phase shift is set to $\Delta=\pi/4$, which forms the channel matrix:
$
H_{1}=\begin{bmatrix}
1 & e^{j\pi/4} \\
e^{j\pi/4} & 1
\end{bmatrix}.
$
Second, the phase shift is tuned to $\Delta=\pi/2$, the optimal condition to minimize crosstalk penalty as reported in [15], yielding the channel matrix:
$
H_{2}=\begin{bmatrix}
1 & e^{j\pi/2} \\
e^{j\pi/2} & 1
\end{bmatrix}.
$
We evaluate the information rate (IR) performance of both MIMO channels under phase noise impairments for the OFC and LD2 configurations. For fair comparison, the optical linewidth parameter is uniformly set to 2.1 MHz, which is extracted from the measurement results in [21], and the relative time delay of the OFC system is fixed at 0.1 ns. The corresponding phase-noise-free AWGN MIMO channels are also simulated as benchmark references, and 64QAM modulation is adopted for all test cases. The simulated IR curves varying with SNR are illustrated in Fig. 13. At an SNR of 20 dB, for the $H_{1}$ channel, the OFC configuration achieves an IR of 11.03 bit/sym, while the LD2 configuration only reaches 10.90 bit/sym. Relative to the ideal AWGN MIMO benchmark, the phase-noise-induced IR losses are 0.13 bit/sym and 0.25 bit/sym for OFC and LD2, respectively. For the optimized $H_{2}$ channel with $\Delta=\pi/2$, the OFC configuration attains an IR of 11.48 bit/sym and the LD2 counterpart achieves 11.41 bit/sym; the corresponding IR degradations compared with the ideal AWGN reference are 0.12 bit/sym and 0.19 bit/sym. It is worth noting that the maximum achievable IR of a 64QAM-modulated SISO channel is limited to only 6.00 bit/sym. As clearly demonstrated by all curves in Fig. 13 across the full SNR range from 14 dB to 26 dB, the $2\times2$ LOS-MIMO architecture provides a substantial capacity gain over conventional SISO transmission. In addition, the OFC scheme consistently outperforms the LD2 scheme under the same channel matrix, and the $H_{2}$ optimal phase configuration always delivers a higher IR than the $H_{1}$ channel for both system configurations.

\begin{figure}[!t]
    \centering
    \includegraphics[width=\linewidth]{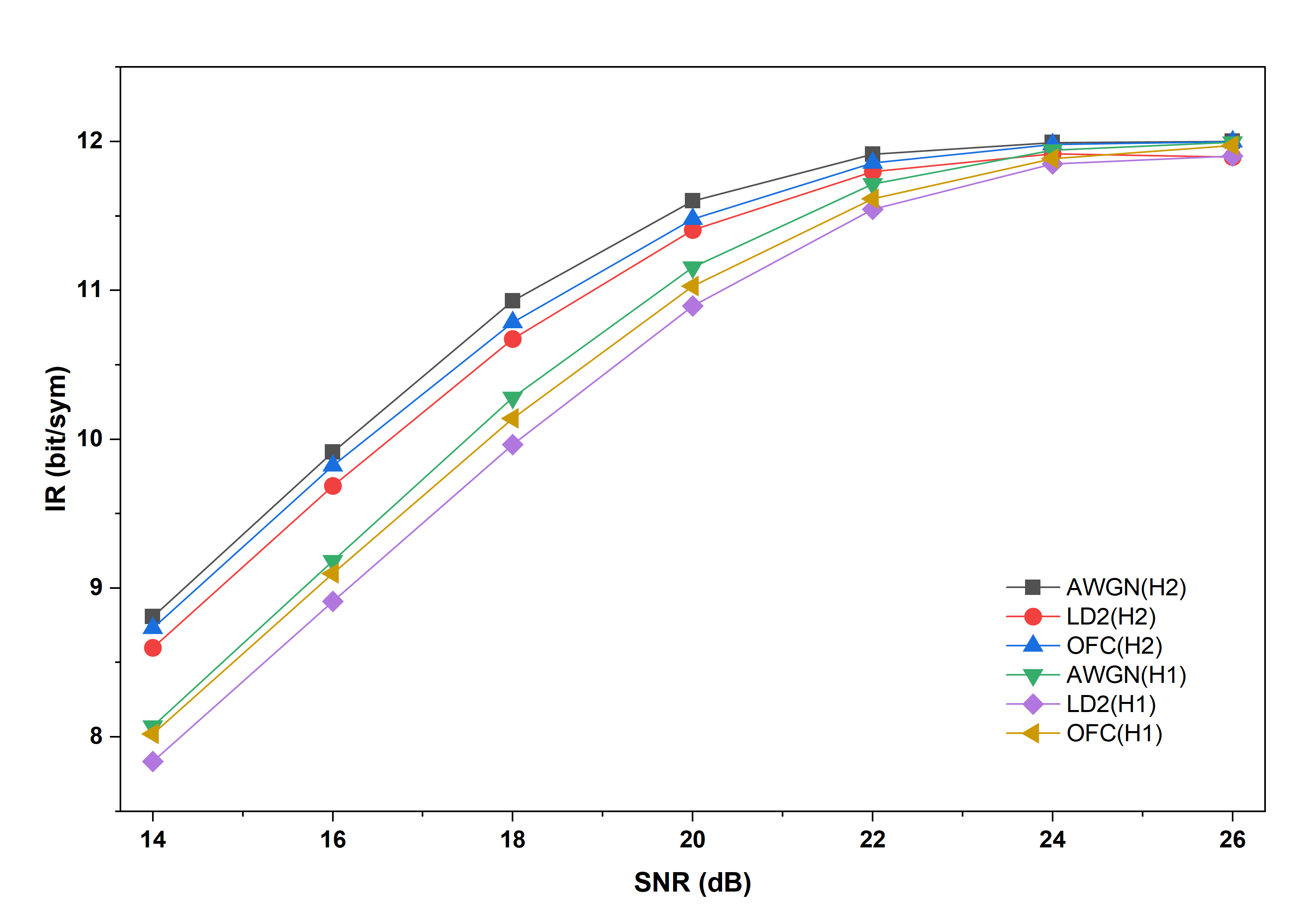}
    \caption{Calculated IR versus SNR for $2\times 2$ LOS-MIMO channels with phase shift parameters $\Delta=\pi/4$ and $\Delta=\pi/2$. AWGN MIMO channels are shown for reference.}
    \label{fig:mimo}
\end{figure}

\section{Conclusion}
This paper investigated the IR of fiber–wireless communication systems employing photonic generation of RF signals. Two primary system configurations were considered: (1) systems utilizing two free-running lasers and (2) systems based on an OFC. Comprehensive system models were established for both configurations. A novel numerical algorithm based on the trapezoidal-rule quadrature was proposed to efficiently calculate the IR for these systems. Numerical simulations validated the effectiveness of the proposed approach and provided useful insights into the design and optimization of fiber–wireless communication systems employing photonic RF signal generation.
\appendices
\section{Derivation of $p(\xi_k|\xi_{k-1})$}
First, we define
\begin{align}
\xi_k &= \xi(kT_0) = \beta(kT_0)-\beta(kT_0-\tau),\notag\\
\xi_{k-1} &= \xi((k-1)T_0) = \beta((k-1)T_0)-\beta((k-1)T_0-\tau).
\end{align}
We then derive the transition PDF $p(\xi_k|\xi_{k-1})$ by considering two distinct cases depending on the relationship between the symbol interval $T_0$ and the time delay $\tau$.\par
\indent When $\tau\leq T_0$, we have
\begin{align}
(k-1)T_0-\tau \leq (k-1)T_0  \leq kT_0-\tau \leq kT_0  
\end{align}
Thus, $\xi_k$ and $\xi_{k-1}$ are composed of disjoint increments of the Wiener process $\beta(t)$ and are therefore independent. Consequently, the conditional PDF reduces to the marginal PDF $p(\xi|\xi_{k-1})=p(\xi_k)$.\\
\indent Next, we consider the case when $\tau>T_0$. In this case, we have the following time ordering
\begin{align}
(k-1)T_0-\tau < kT_0-\tau < (k-1)T_0 < kT_0.
\end{align}
Therefore, $\xi_k$ and $\xi_{k-1}$ can be decomposed as
\begin{align}
\xi_{k-1} &= \beta((k-1)T_0)-\beta(kT_0-\tau) \notag\\
&+ \beta(kT_0-\tau)-\beta((k-1)T_0-\tau) \notag\\
&= b+a,\notag\\
\xi_k &= \beta(kT_0)-\beta((k-1)T_0) \notag\\
&+ \beta((k-1)T_0)-\beta(kT_0-\tau) \notag\\
&=c+b,
\end{align}
where
\begin{align}
a &= \beta(kT_0-\tau) - \beta((k-1)T_0-\tau)\notag\\
b &= \beta((k-1)T_0)-\beta(kT_0-\tau)\notag\\
c &= \beta(kT_0)-\beta((k-1)T_0)
\end{align}

Since $a,b,c$ are disjoint increments of the Wiener process, they are mutually independent zero-mean Gaussian random variables
\begin{align*}
a &\sim \mathcal{N}\bigl(0,\sigma_a^2=2\pi B_l T_0\bigr), \\
b &\sim \mathcal{N}\bigl(0,\sigma_b^2=2\pi B_l (\tau-T_0)\bigr), \\
c &\sim \mathcal{N}\bigl(0,\sigma_c^2=2\pi B_l T_0\bigr).
\end{align*}
Therefore, the PDF of $\xi_{k-1}$ is given by
\begin{align}
p(\xi_{k-1}) = \frac{1}{\sqrt{2\pi (\sigma_a^2 + \sigma_b^2)}} 
\exp\left( -\frac{\xi_{k-1}^2}{2(\sigma_a^2 + \sigma_b^2)} \right). 
\end{align}

Since
\begin{align*}
\operatorname{Var}(\xi_{k-1}) &= \operatorname{Var}(a+b) = \sigma_a^2 + \sigma_b^2, \\
\operatorname{Var}(\xi_k) &= \operatorname{Var}(b+c) = \sigma_b^2 + \sigma_c^2, \\
\operatorname{Cov}(\xi_{k-1},\xi_k) &= \operatorname{Cov}(a+b,b+c) = \sigma_b^2,
\end{align*}
the PDF of $[\xi_{k-1},\xi_k]^T$ can be given by
\begin{align}
\begin{bmatrix}
\xi_{k-1}\\
\xi_k    
\end{bmatrix}
\sim
\mathcal{N}\bigg(
\begin{bmatrix}
0\\
0    
\end{bmatrix},
\begin{bmatrix}
\sigma_a^2+\sigma_b^2,\sigma_b^2\\
\sigma_b^2, \sigma_b^2+\sigma_a^2    
\end{bmatrix}
\bigg)  
\end{align}
Therefore, the joint PDF of $\xi_{k-1}$ and $\xi_k$ is given by
\begin{align}
&p(\xi_k, \xi_{k-1}) = 
\frac{1}{2\pi \sqrt{\sigma_a^2\sigma_b^2 + \sigma_a^2\sigma_c^2 + \sigma_b^2\sigma_c^2}}\times \notag\\
&\exp\left[
-\frac{ (\sigma_a^2 + \sigma_b^2)\xi_k^2 - 2\sigma_b^2 \xi_k \xi_{k-1} + (\sigma_b^2 + \sigma_c^2)\xi_{k-1}^2 }
{2\left( \sigma_a^2\sigma_b^2 + \sigma_a^2\sigma_c^2 + \sigma_b^2\sigma_c^2 \right)}
\right]    
\end{align}

Finally, the conditional PDF $p(\xi_k \mid \xi_{k-1})$ can be expressed as
\begin{align}
p(\xi_k \mid \xi_{k-1}) &= p(\xi_k,\xi_{k-1})/p(\xi_{k-1}) \notag\\
&= \frac{1}{\sqrt{2\pi \sigma_t^2}}
\exp\left(
-\frac{1}{2\sigma_t^2}
\left(\xi_k - \kappa\xi_{k-1}\right)^2
\right).
\end{align}
where
\begin{align}
\sigma_t^2 &= \frac{\sigma_a^2 \sigma_b^2 + \sigma_a^2 \sigma_c^2 + \sigma_b^2 \sigma_c^2}{\sigma_a^2 + \sigma_b^2}=2\pi B_l T_0\big(2-\frac{T_0}{\tau}\big) \notag\\
\kappa &= \frac{\sigma_b^2}{\sigma_a^2+\sigma_b^2} = 1-\frac{T_0}{\tau}
\end{align}

\section{Derivation of $p(\boldsymbol{y}_k|\boldsymbol{x}_k)$ for the memoryless OFC configuration}
This derivation is essential for calculating the IR when the OFC configuration is memoryless, i.e., when $\tau<T_0$.\\
\indent Let $I_l(\cdot)$ denote the modified Bessel function of the first kind of order $l$, Then, the conditional PDF $p(\boldsymbol{y}_k|\boldsymbol{x}_k,\xi_k)$ can be expressed as
\begin{align}
&p(\boldsymbol{y}_k|\boldsymbol{x}_k,\xi_k)\notag\\
&=\frac{1}{(2\pi\sigma_n^2)^B}\exp\Big( -\frac{|\!|\boldsymbol{y}_k|\!|^2+|\!| H\boldsymbol{\boldsymbol{x}_k} |\!|^2}{2\sigma_n^2} \Big)\exp\Big( \frac{|w|}{\sigma_n^2}\cos(\xi_k-\phi) \Big)  \notag\\
&= \frac{1}{(2\pi\sigma_n^2)^B}\exp\Big( -\frac{|\!|\boldsymbol{y}_k|\!|^2+|\!| H\boldsymbol{\boldsymbol{x}_k} |\!|^2}{2\sigma_n^2} \Big)\sum_{l=-\infty}^\infty I_l(\frac{|w|}{\sigma_n^2})e^{jl(\xi_k-\phi)}
\end{align}
where
\begin{align}
w &= \boldsymbol{x}_k^{\dagger}H^{\dagger}\boldsymbol{y}_k, \notag\\
\phi &= \arg(w).
\end{align}
The wrapped normal prior $p(\xi_k)$ can be written as
\begin{align}
p(\xi_k) = \frac{1}{2\pi}\sum_{m=-\infty}^\infty e^{-m^2\sigma_{\xi}^2(\tau)/2}e^{jm\xi_k}    
\end{align}
By marginalizing over the latent phase $\xi_k \in [-\pi,\pi]$, the marginal PDF $p(\boldsymbol{y}_k|\boldsymbol{x}_k)$ is derived as
\begin{align}
&p(\boldsymbol{y}_k|\boldsymbol{x}_k) \notag\\
&=\int_{-\pi}^{\pi} p(\boldsymbol{y}_k \mid \boldsymbol{x}_k,\xi_k)\,p(\xi_k)d\xi_k \notag\\
&= \frac{1}{(2\pi\sigma_n^2)^B}\exp\Big( -\frac{|\!|\boldsymbol{y}_k|\!|^2+|\!| H\boldsymbol{\boldsymbol{x}_k} |\!|^2}{2\sigma_n^2} \Big) \notag\\
&\times
\Bigg[
I_0\!\left(
\frac{|w|}{\sigma_n^2}
\right)
+
2\sum_{m=1}^{\infty}
I_m\!\left(
\frac{|w|}{\sigma_n^2}
\right)
\exp\!\left(
-\frac{m^2\sigma_{\xi}^2(\tau)}{2}
\right)
\cos(m\phi)
\Bigg]
\end{align}

\section{Error analysis}
This appendix provides an approximation error analysis to compare the numerical accuracy of the conventional AC algorithm and the proposed trapezoidal rule-based algorithm for IR calculation.\\
\indent The AC algorithm proposed in \cite{barletta2012information} can be interpreted as a midpoint quadrature method for IR calculation. To elaborate on this inherent numerical integration characteristic, we reformulate the core update formula of $\beta_k$ in the AC algorithm (the update formula of $\alpha_k$ yields identical conclusions). The recursive expression of $\beta_k(\varphi_k)$ is given by
\begin{align}
{\beta_k}({\varphi_k})
&=p({{\boldsymbol{y}}_k}|{\varphi_k})\int_{-\pi }^\pi p({\varphi_k}|{\varphi_{k-1}}){\beta_{k-1}}({\varphi_{k-1}})d{\varphi_{k-1}}\notag\\
&= p({\boldsymbol{y}_k}|{\varphi _k})\sum_{j = 0}^{T - 1} I_j
\end{align}
where the sub-integral term $I_j$ corresponding to the $j$-th subinterval is defined as
\begin{align}
I_j &= \int_{jh}^{(j+1)h} p({\varphi _k}|{\varphi _{k - 1}})\beta_{k - 1}({\varphi_{k-1}})d{\varphi _{k - 1}}.
\end{align}
\indent Applying the midpoint quadrature for integral, each $I_j$ is then approximated by assuming $\beta_{k-1}(\varphi_{k-1})$ is constant over the sub-interval and equal to its value at the midpoint $\beta_{k-1}((j+1/2)h)$. The approximate sub-integral $I_j$ is formulated as
\begin{align}
K_j = {\beta_{k-1}}((j+1/2)h)\int_{jh}^{(j+1)h}p({\varphi _k}|{\varphi_{k-1}})d{\varphi_{k-1}}.
\end{align}
Substituting this approximation into the recursive formular $\beta_k$ yields
\begin{align}
&{\beta_k}({\varphi_k=(i+1/2)h}) \notag\\
&\approx p(\boldsymbol{y}_k|(i+1/2)h)\sum_{j = 0}^{T - 1} K_j \notag\\
&= p(\boldsymbol{y}_k|(i+1/2)h)\sum_{j = 0}^{T - 1}\beta_{k-1}((j+1/2)h) \notag\\
&\times \int_{jh}^{(j+1)h}p((i+1/2)h|{\varphi _{k - 1}})d{\varphi_{k-1}} \notag\\
&= p(\boldsymbol{y}_k|(i+1/2)h)\sum_{j=0}^{T-1}\beta_{k-1}((j+1/2)h)P_{i,j}.
\end{align}
where the transition probability $P_{i,j}$ is given by
\begin{align}
{P_{i,j}} &= \int_{jh}^{(j+1)h}p((i+1/2)h|{\varphi _{k - 1}})d{\varphi _{k - 1}} \notag\\
&= \sum_{l = -\infty}^\infty Q\left(\frac{(j-i-1/2)h - 2\pi l}{\sigma _\varphi}\right) \notag\\
&-\sum_{l = -\infty}^\infty Q\left(\frac{(j - i+1/2)h - 2\pi l}{\sigma _\varphi}\right),   
\end{align}
where $Q(x)$ is the Gaussian Q function.
The above formula is consistent with Eq. (6) in \cite{barletta2012information}.\\
\indent Based on the theorem 2 from Chapter 6.1 of \cite{NA}, the approximation error $E_j$ between the orginal sub-integral $I_j$ and its mid-point quadrature $K_j$ can be quantified as
\begin{align}
{E_j} &= {I_j} - {K_j} \notag\\
&= \int_{jh}^{(j+1)h}{\beta'}_{k-1}(\xi ({\varphi_{k-1}}))({\varphi_{k-1}}-(j+1/2)h) \notag\\
&\quad \times p({\varphi_k}|{\varphi_{k-1}})d{\varphi_{k-1}}.
\end{align}
where ${\beta'}$ is the first-order derivative of ${\beta}$ and $\xi \in (jh,(j+1)h)$ denotes an arbitrary point within the sub-interval. Applying the first mean value theorem for definite integrals, we obtain 
\begin{align}\label{err_AC}
{E_j}
&= \int_{jh}^{(j+1/2)h}{\beta'}_{k - 1}(\xi ({\varphi _{k - 1}}))({\varphi _{k - 1}} - (j+1/2)h) \notag\\
&\quad p({\varphi _k}|{\varphi _{k - 1}})d{\varphi _{k - 1}} \notag\\
&\quad + \int_{(j+1/2)h}^{(j+1)h}{\beta '}_{k - 1}(\xi ({\varphi _{k - 1}}))({\varphi _{k - 1}} - (j+1/2)h) \notag\\
&\quad p({\varphi _k}|{\varphi _{k - 1}})d{\varphi _{k - 1}} \notag\\
&= {\beta'}_{k - 1}(a)p({\varphi _k}|b)\int_{jh}^{(j+1/2)h}({\varphi _{k - 1}} - (j+1/2)h)d{\varphi _{k - 1}} \notag\\
&\quad + {\beta'}_{k - 1}(\gamma )p(\delta )\int_{(j+1/2)h}^{(j+1)h}({\varphi _{k - 1}} - (j+1/2)h)d{\varphi _{k - 1}} \notag\\
&= \frac{{\beta '}_{k - 1}(\gamma )p(\delta ) - {\beta '}_{k - 1}(a)p({\varphi _k}|b)}{8}h^2. 
\end{align}
where $a,b\in(jh,(j+1/2)h)$ and $\gamma,\delta\in((j+1/2)h,(j+1)h)$.
It is evident from Eq. (\ref{err_AC}) that the error of the AC algorithm obeys the order of $\mathcal{O}(h^2)$. 
\par
Subsequently, the approximation error of the proposed trapezoidal rule-based algorithm is analyzed following the same framework. According to the theorem 2 in Chapter 6.1 given in \cite{NA}, the approximation error of the $j$-th sub-integral in the proposed algorithm is given by
\begin{align}
{E_j} &= {I_j} - {K_j} \notag\\
&= \int_{jh}^{(j + 1)h} \frac{{\beta ''}_{k - 1}(\xi ({\varphi _{k - 1}}))}{2}({\varphi _{k - 1}} - jh)({\varphi _{k - 1}} - (j + 1)h) \notag\\
&\quad \times p({\varphi _k}|{\varphi _{k - 1}})d{\varphi _{k - 1}}
\end{align}
Similarly, utilizing the first mean value theorem for definite integrals, we can obtain
\begin{align}
{E_j} &= \int_{jh}^{(j+1)h} \frac{{\beta''}_{k - 1}(\xi ({\varphi _{k-1}}))}{2}({\varphi_{k-1}}-jh) \nonumber\\
&\quad \times ({\varphi _{k - 1}} - (j + 1)h)p({\varphi _k}|{\varphi _{k - 1}})d{\varphi _{k - 1}} \nonumber\\
&= \frac{{\beta''}_{k-1}(a)}{2}p({\varphi _k}|b)\int_{jh}^{(j+1)h}({\varphi _{k - 1}} - jh) \nonumber\\
&\quad \times ({\varphi _{k - 1}} - (j + 1)h)d{\varphi _{k - 1}} \nonumber\\
&= -\frac{{\beta''}_{k - 1}(a)}{12}p({\varphi _k}|b )h^3.
\end{align}
where $a,b \in (jh, (j+1)h)$.
This result demonstrates that the approximation error of the proposed algorithm scales with $\mathcal{O}(h^3)$. 
\par
The comparative error analysis verifies the superior numerical accuracy of the proposed algorithm over the conventional AC method. Benefiting from the higher-order error convergence $\mathcal{O}(h^3)$ versus $\mathcal{O}(h^2)$, the proposed trapezoidal rule-based scheme achieves smaller approximation errors, faster convergence speed, and higher calculation precision with fewer discrete subintervals. This theoretical conclusion well explains the outstanding numerical performance observed in simulations, particularly for accurate IR estimation in fiber-wireless communication systems.

\ifCLASSOPTIONcaptionsoff
  \newpage
\fi



\bibliographystyle{IEEEtran}
\bibliography{Bib/IEEEabrv,Bib/mylib}

\end{document}